\documentclass[%
 aip,
 jmp,%
 amsmath,amssymb,
 reprint,%
]{revtex4-1}

\usepackage{graphicx}
\usepackage{dcolumn}
\usepackage{bm}
\usepackage{mathtools}
\usepackage{hyperref}
\usepackage{mathrsfs}

\hypersetup{
    colorlinks,
    citecolor=black,
    filecolor=black,
    linkcolor=black,
    urlcolor=black
}

\begin{document}


\title[Field Theory for Magnetic Monopoles in  (Square, Artificial) Spin Ice]{Field Theory for Magnetic Monopoles in  (Square, Artificial) Spin Ice}

\author{Cristiano Nisoli}

\affiliation{ 
Theoretical Division, Los Alamos National Laboratory, Los Alamos, NM, 87545, USA
}%


\date{\today}

\begin{abstract}
Proceeding from the more general to the more concrete, we propose an equilibrium field theory describing spin ice systems in terms of topological charges and magnetic monopoles. We show that for a spin ice on a graph,  the entropic interaction in a Gaussian approximation is the inverse of the graph Laplacian matrix, while the screening function for external charges is the inverse of the screened laplacian. We particularize the treatment to  square and pyrochlore ice. For square ice we highlight the gauge-free duality between direct and perpendicular  structure in terms of symmetry between charges and currents,  typical of magnetic fragmentation in a two-dimensional setting. We derive structure factors, correlations, correlation lengths, and susceptibilities  for spins, topological charges, and currents. We show that the divergence of the correlation length at low temperature is exponential and inversely proportional to the mean square charge.  While in three dimension real and entropic interactions among monopoles are both 3D-Coulomb, in two dimension the former is a 3D-Coulomb and the latter 2D-Coulomb, or logarithmic, leading to weak singularities in correspondence of the pinch points and destroying charge screening. This suggests that  the  monopole plasma of square ice is a magnetic charge  insulator.
\end{abstract}

\maketitle

\tableofcontents

\section{Introduction}

Since the Bernal-Fowler ice rule~\cite{bernal1933theory} was invoked by Pauling~\cite{Pauling1935} to explain the zero-point entropy of water ice~\cite{giauque1933molecular,giauque1936entropy} the concept has come to describe a variety of other materials, such as pyroclore rare-earth spin ice antiferromagnets~\cite{Ramirez1999,den2000dipolar,Bramwell2001}, artificial magnetic spin ice antiferromagnets~\cite{tanaka2006magnetic,Wang2006,Nisoli2013colloquium,heyderman2013artificial,skjaervo2019advances}, or artificial particle-based ices~\cite{ortiz2019colloquium}.
Often, but not always, in these materials frustration impedes ordering among binary degrees of freedom even at low temperature and leads to degenerate states of  constrained disorder, or {\it ice-manifolds} of interesting topological properties. 

Indeed the ice rule is a topological concept related to the local minimization of a topological charge~\cite{nisoli2014dumping,nisoli2018unexpected}. As such it has a wide applicability, and artificial spin ice materials are being designed~\cite{nisoli2017deliberate} for a variety of emergent behaviors not necessarily found in natural magnets~\cite{nisoli2017deliberate,nisoli2018frustration,Zhang2013,gilbert2016emergent,gilbert2014emergent,lao2018classical}. The breakdown of the ice manifold is  associated with the emergence of topological excitations~\cite{ryzhkin2005magnetic,Castelnovo2008} that, depending on the geometry and local degeneracy of the system, can be deconfined. Further, in magnetic materials these topological charges are also {\it magnetic charges} often deconfinable as {\em monopoles}~\cite{ryzhkin2005magnetic,Castelnovo2008,morris2009dirac,Giblin2011, ladak2011direct,Mengotti2010}. They interact via a Coulomb law, are sources and sinks of the $H$ field, and can pin superconductive vortices in spin ice/superconductors heterostructures~\cite{wang2018switchable}. They might exert  a localized {\it and} mobile magnetic proximity effect~\cite{scharf2017magnetic} in heterostructures that interface two-dimensional spin ices  to transition metal dichalcogenides  or Dirac materials. Finally, in artificial realizations, these topological objects can also be read and written~\cite{wang2016rewritable,gartside2018realization}. They might therefore function as binary, mobile information carriers, opening new perspectives to spintronics. Artificial versions of these materials are being explored for neuromorphic  computation~\cite{caravelli2018computation,arava2018computational}. 
 
In non-magnetic spin ices---such as particle-based ones~\cite{ortiz2019colloquium}, which can be made of confined colloids~\cite{Libal2006,ortiz2016engineering,loehr2016defect,libal2018ice}, superconducting vortices~\cite{libal2009,Latimer2013,Trastoy2013freezing,wang2018switchable}, skyrmions in magnets~\cite{ma2016emergent} or liquid crystals~\cite{duzgun2019artificial,duzgun2019commensurate}---the mutual interaction among monopoles differs from a Coulomb law~\cite{libal2017,nisoli2018unexpected}, though, because they are  topological charges, they  always interact at least entropically in a thermal ensemble~\cite{libal2018ice}, as we shall see. 

Here we treat  the ice manifold and its excitations on general grounds, and we employ topological/magnetic charges as elementary degrees of freedom. Various interesting effective field theories of the ice manifold in pyrochlore, often called {\it Coulomb phase}, have been developed, generally to obtain dipolar correlations via a coarse grained field~\cite{henley2010coulomb,isakov2004dipolar,garanin1999classical,henley2005power}. They also pertain to similar systems, such as the dimer cover problem or generally height models~\cite{henley2010coulomb,henley2011classical,henley1997relaxation}. However, because  there is in general no transition to the spin ice manifold, but only a crossover at a temperature set by the energy scale of the monopoles, the criticality of the ice manifold cannot be reached: monopoles, however diluted, will always be present at any non-zero $T$, their density setting a correlation length. Close to zero temperature, the model breaks down and further neighbor interactions are expected to induce  ordering~\cite{melko2001long}, which might  however be prevented by glassiness~\cite{castelnovo2010thermal}. 

Therefore, while the topological protection of the ice manifold is prima facie enticing, the exotic behaviors of spin ices proceed  not so much from said topological structure, but rather from how it is {\it broken}, e.g.\ via fractionalization into monopoles~\cite{ryzhkin2005magnetic,Castelnovo2008,castelnovo2012spin}, and how much of it is instead retained,  e.g.\ via spin fragmentation~\cite{brooks2014magnetic,canals2016fragmentation,petit2016observation}. So far, the phenomenology of monopole kinetics has been attacked either numerically or via phenomenological theories (generally by adapting a Debye-H\"uckel/Bjerrum approach), or both~\cite{ryzhkin2005magnetic,Giblin2011,castelnovo2011debye,castelnovo2010thermal,mol2009magnetic,farhan2019emergent,jaubert2011magnetic,Nascimento2012,mol2010conditions,dusad2019magnetic,klyuev2017statistics,kirschner2018proposal,castelnovo2010coulomb}. This proved most useful in describing the specificity of the experimental reality, including out of equilibrium scenarios.

We shall instead try a more unifying framework, which is certainly exact in its formulation, but not in its approximated deductions. 
We will begin with a graph theory approach. This  choice is motivated by  various reasons. 

Firstly, while graphs do have a metric structure, they are not embedded in a linear algebraic structure, each of them essentially describing a topological class of various and different geometric realizations. This implies that  many topological concepts related to these spin ice materials, currently disseminated through many  works pertaining to a variety of systems and beginning as early as the late nineties, become natural and unified when treated with tools from graph theory. Indeed we will show that at least in a pure spin ice Hamiltonian, and at least in a high temperature limit (but as we shall see even at lower temperatures), informations about correlations, screening, et cetera are completely obtainable from the graph spectral analysis, a subject widely studied  by mathematicians~\cite{brouwer2011spectra}.

Secondly,  in the spirit of guiding artificial realizations~\cite{Nisoli2013colloquium,heyderman2013artificial,skjaervo2019advances,ortiz2019colloquium,nisoli2018frustration}, while we are more directly motivated by recent realizations of modified square ice~\cite{Moller2006,perrin2016extensive,ostman2018interaction,farhan2019emergent} that are also thermally active~\cite{Kapaklis2012,farhan2013direct,kapaklis2014thermal,arnalds2012thermalized,anghinolfi2015thermodynamic}, we aim  to a  generality that can suggest conceptualizations applicable to artificial materials that do not yet exist. Instead of finding correlations in a spin ice of a certain topology, we could begin from observables and then realize graph-based spin ices whose spectral properties make them behave in desired ways. This could apply to the many-body physics of reprogrammable~\cite{chern2017magnetotransport}  electrical circuits of connected spin ices, going beyond  the recently explored Kagome~\cite{Branford2012,PhysRevB.95.060405}, to realize potentially useful devices. A graph treatment can conceptualize problems of  frustrated social interactions and exchanges~\cite{mahault2017emergent}, while the ice rule on a graph can also describe compatibility of soft modes in artificial mechanical meta-materials~\cite{meeussen2019topological,coulais2016combinatorial}.  

Thirdly, there is a promising line of works on new, exotic topologies~\cite{Morrison2013,Chern2013,nisoli2018frustration,gilbert2014emergent, gilbert2016emergent,lao2018classical,nisoli2018frustration}, including finite size systems~\cite{Li2010}. Especially interesting works on Penrose spin ices, based on finite size quasicrystals~\cite{bhat2013controlled,bhat2014non,brajuskovic2018observation,bhat2014ferromagnetic} which still defy complete understanding, and for which concepts explained in the next section might possibly prove useful.  While the Hamiltonian in Eq.~(\ref{Hs}) is likely too simplified to describe systems that lack degeneracy at the vertex level, it is nonetheless a good starting point, and we show in the section on square ice how ordering terms can be introduced by favoring topological currents. 
  
The article is structured as follows. In the first section we will distill in mathematical form general ideas---such as what is a spin ice, an ice manifold, a Coulomb phase, its degeneracy, et cetera---on a general graph and show how its many-body properties can be deduced from the spectrum of the graph, at least at high $T$. In the following section we  particularize such descriptions to  the recently realized square degenerate ice~\cite{Moller2006,perrin2016extensive,ostman2018interaction,farhan2019emergent}  while drawing comparisons with natural pyrochlore magnets~\cite{Ramirez1999,den2000dipolar,Bramwell2001}. 

The readers only interested in those more physically grounded cases can jump directly there after reading only the initial part of the next section. Indeed we have kept the second part as much as possible independent from the first. There, and indeed elsewhere, we do not shy away from  repetitions when we think they might be convenient to the reader.

\section{Graph Spin Ice }

We consider the most general case of a spin ice on a connected, undirected, simple graph $G$ ~\cite{west2001introduction}. First we need to assign on it a notion of binary spins. That corresponds to a notion of directionality on the graph which allows us to talk of a {\it spin phase space} for $G$, or the set of all directed graphs that can be built on $G$. From that we can define an {\it ice manifold} as the subset of the spin phase space of $G$ that minimizes a proper notion of topological charge. 

\subsection{Spins on a Graph}

 To begin, consider an undirected, simple graph $G$ of a number $N_l$ of edges labeled by $l$, connecting a number $N_v$ of vertices labeled by $v$ and of various degree of coordination $z_v$. We call $\{ vv'\}$ an edge $l$ among vertices $v, v'$. For such graph, the {\em adjacency matrix}~\cite{west2001introduction} is the matrix  $A_{vv'}$, such that  $A_{vv'}=1$ if $v, v'$ are connected and $A_{vv'}=0$ otherwise. It thus contains all the topological information of the graph. Obviously, $A_{vv'}$ is symmetric and $z_v=\sum_{v'} A_{vv'}$.

We can define binary variables or Ising spins $S_l$ on each edge $l$ such that it ``points'' either toward $v$ or $v'$, as in Fig.~1. That can be expressed via an antisymmetric matrix $S_{vv'}$ such that $S_{vv'}=0$ if $v$ and $v'$ {\it do not} share an edge, $S_{vv'}=1$ if they do and the spin points toward $v'$, and $S_{v'v}=-1$  if they do and the spin points toward $v$.  

In the language of graph theory, $A_{vv'}$ defines an undirected graph, while an Ising spin structure $S_{vv'}$ defines a directed graph. Given an undirected graph $G$, we call ${\mathscr S}$, or the {\it spin phase space} of $G$, the set of  the $2^L$ directed graphs that can be specified (via $S_{vv'}$) on it. As such, each $S$ matrix  can be related to the non-symmetric adjacency matrix of the corresponding directed graph $A^{\mathrm{dir}}_{vv'}$, whose elements have value one if and only if $vv'$ are connected by an edge pointing toward $v'$ (on simple graphs). Then $S$ is the anti-symmetrization of $A^{\mathrm{dir}}$, or $S_{vv'}=A^{\mathrm{dir}}_{vv'}-A^{\mathrm{dir}}_{v'v}$ (and of course  $A_{vv'}= A^{\mathrm{dir}}_{vv'}+A^{\mathrm{dir}}_{v'v}$).

\subsubsection{Ice Manifold}

Having defined the spin phase space ${\mathscr S}$ as the set of the directed graphs realizable on a certain undirected graph $G$, we can define an {\it ice manifold} as a proper subset of ${\mathscr S}$. Given $S_{vv'}$, for each vertex $v$ we can then define its topological charge as the vector $Q_v$  defined as the difference between the edges pointing in and out of $v$, or
\begin{equation}
Q_v[S]=\sum_{v' } S_{v'v}.
\label{charge}
\end{equation}
$Q_v$ is thus the difference between {\it indegrees} and {\it outdegrees} of the directed graph that corresponds to a particular spin configuration on $G$. $Q_v$ can have the values $z_v, z_v-2, \dots, 2-z_v, -z_v$, and thus only vertices of even coordination can have zero charge. To relate  Eq.~(\ref{charge}) to more familiar pictures~\cite{henley2010coulomb}, we can introduce the generalized divergence operator as
\begin{equation}
\mathrm{div}[S]_v=\sum_{vv'}S_{vv'}
\label{div}
\end{equation}
from which we immediately have
\begin{equation}
Q_v[S]=-\mathrm{div}[S]_v.
\label{charge2}
\end{equation}
\begin{figure}[t!]
\includegraphics[width=.5\columnwidth]{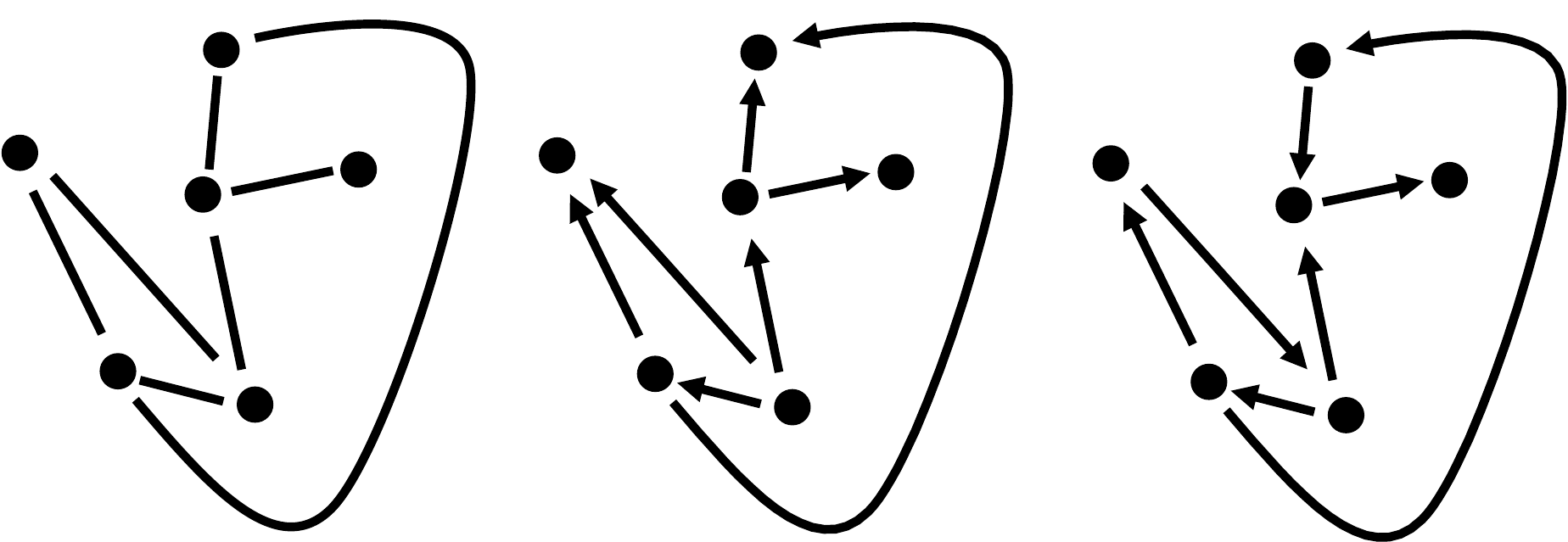}
\caption{Left: undirected graph. Center: directed graph. Right: directed graph obeying the ice rule.  }
\label{fig1_0}
\end{figure}

A directed graph for which $|Q_v|$ is minimal on each vertex, i.e\ has zero charge on all vertices of even coordination and $\pm1$ charge on all vertices of odd coordination, is said to {\it obey the ice-rule}.  Then, given a graph $G$, we call the {\it ice-manifold} of $G$ the subsets ${\mathscr I} \subset {\mathscr S}$ of its spin phase space ${\mathscr S}$  made of all directed graphs that obey the ice rule. 

Depending on the topology of the graph it is not obvious  if and when the cardinality of the ice manifold should scale exponentially with the size of the graph, thus leading to a non-zero density of entropy, though an argument \`a la Pauling~\cite{Pauling1935} would suggest so. Indeed, the one dimensional ferromagnetic Ising model can be mapped into a spin ice on a path graph whose ice manifold has cardinality two, regardless of the number of vertices. Here we will consider cases of non-empty and extensively degenerate ice manifolds.  

\subsubsection{Coulomb Phases}

The concept of a Coulomb phase appeared first in gauge field theories~\cite{fradkin2013field} and was introduced in pyrochlore spin ice by C. Henley~\cite{henley2010coulomb}. In simple terms, it corresponds to a disordered spin texture that can be coarse-grained to a solenoidal magnetization field. It can be considered a case of classical topological order~\cite{henley2011classical,lamberty2013classical,castelnovo2012spin} where no order parameter exists but instead the disordered states are labeled by a field which expresses a constraint over the disorder. We generalize it to a graph by expressing it without coarse graining or gauge theories. 

We say that two spin assignations  $S, S' \in {\mathscr S}$ are {\it charge equivalent} if and only if their difference $S_{vv'}-S'_{vv'}$ has {\it zero} topological charge on every vertex. It is immediate to show that change-equivalence is an equivalence relation and thus induces a partition on the phase space ${\mathscr S}$. We call each class of equivalence in that partition a Coulomb class. Each Coulomb class is characterized by a distribution of charge $Q_v$ on each vertex $v$.

Trivially speaking: a graph made of only two  vertices connected by one edge has a spin phase space of cardinality $2$, corresponding to the two orientations of its only spin. Its ice manifold coincides with the spin phase space, and it contains $2$ Coulomb classes, each of cardinality $1$. 

The ice manifold of pyrochlore ice is a Coulomb class corresponding to charge $Q_v=0$ everywhere. Indeed, that is true for every graph that has only  vertices of even coordination. Clearly, not all ice manifolds are Coulomb phases. For instance the ice manifold of the bipartite honeycomb graph, so-called kagome ice, is not: for any spin configuration there is always at least one spin, and in fact an extensive number of spins, that can be flipped individually, thus changing charge configurations without violating the ice rule. And yet, the kagome ice-manifold can be partitioned into Coulomb classes. Its  Ice II phase~\cite{Moller2009,Chern2011,Rougemaille2011,chioar2016ground,libal2017} corresponds to two Coulomb classes, each of charge alternating in sign $\pm1$ on nearby vertices. Then, the spontaneous symmetry breaking between the two Coulomb classes drives a second order transition to charge ordering (of the Ising universality).

A Coulomb class  becomes important if it represents a low energy state. In such case it defines a {\em Coulomb phase}.  That is the case of spin ices of even coordination, including therefore pyrochlores, whose low energy state, the ice manifold, is also a Coulomb phase, as we have said. In kagome ice,  each of the two Coulomb classes of the Ice-II phase are a Coulomb phase, which might explain why it is so hard to observe it experimentally~\cite{Rougemaille2011,macdonald2011classical,Zhang2013,drisko2015fepd}, whereas the Ice-I phase was found easily~\cite{tanaka2006magnetic,qi2008direct,Nisoli2010}. 

Indeed, in general a Coulomb class imposes topological constraints on the kinetics within the class: it  prohibits single spin flips and requires collective flips that might be extremely unlikely in a realistic spin dynamics. Thus, if a Coulomb class is a Coulomb phase, all kinetics must happen {\em above} that phase, by breaking the topology of the Coulomb class, that is by changing its defining distribution of topological charges. In pyrochlore ices or in degenerate artificial square ices, where the entire ice manifold is a Coulomb phase, the breakage of topological protection consists in the appearance of monopoles. In  such systems all kinetics is monopole kinetics. In the Ice-II phase of kagome ice~\cite{macdonald2011classical} it consists in the  breaking of the $\pm 1$ charge alternation. 

Because  properties of topological order pertain to the Coulomb classes, and as the latter are defined by charges, we will develop a formalism in which charges are the relevant degrees of freedom and reach an effective free energy for charge distribution, which contains the entropic effect of the corresponding Coulomb class.

\subsubsection{Energy}

Having clarified the notion of ice manifold and Coulomb phase on a general graph, we then need a Hamiltonian that has the ice manifold as the ground state and allows us to deal with breakages of the topological protection in terms of  energy costs. 

The following is suitable:
\begin{equation}
{\cal H}[Q] = \frac{\epsilon}{2}\sum_v Q_v^2 +  \frac{\mu}{2}\sum_{v \ne v'} Q_v V_{vv'}Q_{v'}. 
\label{Hs}
\end{equation}
$\epsilon, \mu>0$ are energies and $V_{vv'}$ is a positive-definite, symmetric matrix of zero diagonal. The first term is a nearest neighbor energy that is minimal on the ice manifold. The second term represents the possible  interaction between charges on different vertices, not necessarily neighboring. In common cases, $\mu$  or $V$ can be chosen such that the ice manifold remains the ground state and its effect can be to lower the energy of submanifold of the ice manifold (see for instance the two inner phases of kagome ice~\cite{Moller2009,Chern2011,libal2017}). 


The corresponding partition function reads
\begin{equation}
Z[H,V]=\sum_{\{S_{vv'}\}} e^{-\beta \left({\cal H}[Q]- \frac{1}{2}\sum_{vv'}  S_{vv'} H_{vv'}\right)},
\label{Z}
\end{equation}
where $\beta=1/T$, $H_{vv'}$ is an antisymmetric matrix modeling an external ``magnetic'' field defined on the edges with respect to the orientation $l=vv'$, so that $H_{vv'}\ne0$ if and only if $v$ and $v'$ share an edge, and $H_{vv'}=-H_{v'v}$. $H$ has dimensions of an energy (the Zeeman energy).

Clearly then
\begin{equation}
\langle S_l\rangle_{H,V}=T\frac{\delta \ln Z[H]}{\delta {H_l}}, ~~\langle S_l S_{l'}\rangle=T^2\frac{\delta^2 \ln Z[H]}{\delta {H_l}\delta {H_{l'}}}
\label{Sav}
\end{equation}
are the one- and two-point correlation functions for the spins. 
%

   
\subsection{Field Theory}

The theory below is continuous in the sense that discrete structures (topological charges) are made continuous and discrete degrees of freedom (spins) are removed. Positions (nodes), however, remain discretized, though in practical applications long wavelength  (or low eigenvalues of the graph spectrum) approximations can be used to return the familiar space-continuum picture.

To produce a field theory we need to remove the discrete variables $S_l$. We do so by a common trick. We  insert in the sum of (\ref{Z}) the tautological  expression {$1=(2\pi)^{-N_v}\prod_v \int   dq_v d\phi_v\exp\left[i \phi_v\left(q_v-Q_v\right) \right]$} and then we sum over the spins. We note that 
\begin{align}
\Omega_H[\phi]&=\sum_{\{S_{vv'} \}}e^{-i\sum_v\phi_vQ_v+\frac{\beta}{2}\sum_{vv'}  S_{vv'} H_{vv'}} \\
&={\prod_{{\langle vv' \rangle}}}^*\sum_{S=\pm1}e^{[-i(\phi_{v'}-\phi_{v})+\beta H_{vv'} ]S}
\end{align}
where ${\langle vv' \rangle}$  nearest neighbors and the $\prod^*$ means that edges are only counted once (so if ${\langle vv' \rangle}$ is counted ${\langle v'v \rangle}$ is not). From it we obtain 
\begin{equation}
\Omega_H[\phi]=2^{N_l}{\prod_{{\langle vv' \rangle}}}^*\cos\left(\nabla_{vv'}\phi + i \beta H_{vv' } \right),
\label{F}
\end{equation}
 where 
\begin{equation}
\nabla_{vv'}\phi = \phi_{v'} - \phi_{v}.
\label{discgrad}
\end{equation}
 is the {\it gradient} of $\phi$, or a matrix $\nabla \phi$, defined on the edges only. Note that if the graph is embedded in a linear space and $\vec{vv'}$ is the  vector pointing from $v$ to $v'$ then $\nabla_{vv'}\phi=\vec{vv'} \cdot \vec \nabla\phi + O(|vv'|^2)$.

 $Z[H]$ from (\ref{Z}) can be rewritten as
\begin{equation}
Z[H]=\int \left[ dq \right] \tilde\Omega_H[q] e^{-\beta{\cal H}[q]}  
\label{Z2}
\end{equation}
%
%
where  $\left[ dq \right]= \prod_v dq_v $, ${\cal H}[q]$ is given by Eq.~(\ref{Hs}), and $\tilde \Omega_H[q]$  is the functional Fourier transform of $\Omega_H[\phi]$, or 
\begin{equation}
\tilde \Omega_H[q]=\frac{1}{(2\pi)^{N_v}}\int [d\phi] \Omega_H[\phi] e^{\sum_v i q_v \phi_v}.
\label{F}
\end{equation}
Note that, clearly, the averages and correlations of $q$ are the same as for $Q$ and therefore, for instance $\langle q_v\rangle= \langle Q_v\rangle$ and also
\begin{align}
\langle \mathrm{div}[S]_v \mathrm{div}[S]_{v'}\rangle=\langle q_v q_{v'}\rangle.
\end{align}

We have thus replaced the binary spin variables $S_l$ defined on the edges with a continuum field $q_v$ defined on the discrete set of vertices,  but we have  gained a term $\tilde \Omega[q]$. It represents a generalized degeneracy or density of state for the charge distribution $q_v$, emergent from the many possible underlying spin ensembles compatible with $q_v$.  Indeed we can call ${\cal S}=\ln \tilde \Omega[q]$ the generalized entropy for the charge distribution $q$. Therefore,  the effective free energy at zero loop~\cite{zinn1996quantum}  for $q_v$ is given by the quadratic part of ${\cal H}[q]-T{\cal S}[q]$. Also, in absence of a field, from (\ref{F}), (\ref{Sav}) and the parity of the functions involved, $\langle q_v\rangle= 0$  for every $v$  and $\langle S_l\rangle=0$ for every $l$, as one would indeed expect from trivial considerations on the model.

To see this formalism from a different angle we can rewrite the partition function as
\begin{equation}
Z[H]=\frac{1}{(2\pi)^{N_v}} \int \left[ dq d\phi \right] e^{-\beta{\cal F}[q,\phi,H]}  
\label{QFT}
\end{equation}
with
\begin{equation}
{{\cal F}[q,\phi,H]} ={\cal H}[q]-iT\sum_v q_v \phi_v+ {\cal F}_H[\phi]
\label{QFT2}
\end{equation}
and
\begin{equation}
{\cal F}_H[\phi]=-T \ln \Omega_H[\phi].
\label{QFT3}
\end{equation}
Equations (\ref{QFT}), (\ref{QFT2}) look familiar in the language of quantum field theory. They correspond to a charge  field $q_v$, for which $i T \phi_v$ acts  as a bosonic field (of ``Lagrangian''  ${\cal F}[\phi]$) mediating an interaction   between charges. Again, the interaction is not real but comes from the underlying binary ensemble from which the charge field is an emergent observable. 
Both pictures come useful in different scenarios, as we shall see. 

From now on we will consider for simplicity now $\mu=0$, or $V_{vv'}=0$: we neglect the monopole interaction. The effect of such interaction will be discussed in the next section in the case of pyrochlores and (quasi) degenerate square ices. Integrating  (\ref{QFT}), over $dq_v$ we obtain 
\begin{equation}
Z[H]=\left(\frac{T}{2\pi \epsilon}\right)^{\frac{N_v}{2}}\int \left[ d\phi \right]  \Omega_H[\phi] \exp{\left(-\frac{T}{2\epsilon} \sum_v \phi_v^2\right)} 
\label{Z3}
\end{equation}
which confines  the entropic field at $\langle \phi^2 \rangle\sim \epsilon/T$. Clearly, at large temperature the system becomes uncorrelated and not surprisingly the field that mediates correlations tends to  zero. 

Taking the high $T$, high $H$ limit  but keeping $H/T$  finite, the Gaussians in (\ref{Z3}) tend to delta functions in $\phi$ and we obtain   
\begin{equation}
 Z[H]=2^{N_l}\prod_l \cosh\left( \beta H_l\right), 
 \label{para}
\end{equation}
 which is the standard ``paramagnetic'' partition function for an uncorrelated system. It leads, via Eq.~(\ref{Sav})  to the familiar magnetization
 \begin{equation}
 \langle S_l\rangle=\tanh(\beta H_l).
 \label{trivialM}
 \end{equation}
 Also, when  $H=0$ we get  from (\ref{para}) the correct entropy per spin at high temperature, or  $s=\ln2$.

From  Eq.~(\ref{Z2}) we can immediately deduce a generalization of Eq.~(\ref{trivialM}) as
  \begin{equation}
 \langle S_{vv'}\rangle=\langle \tanh \left(\beta H_{vv'}-i\nabla_{vv'}\phi \right) \rangle,
 \label{M}
 \end{equation}
which is strongly suggestive. It is thus natural to introduce an entropic $B^e$ field, defined as 
\begin{equation}
B^e_{vv'}=-i \nabla_{vv'}\phi
\label{Be}
\end{equation}
such that
\begin{equation}
\langle S_{vv'}\rangle \simeq  \beta H_{vv'} +   \langle  B^e_{vv'} \rangle
\label{M2}
\end{equation}
when $ \langle S_{vv'}\rangle$ is small. 

\subsection{High $T$ Approximation}
\subsubsection{Free Energy}

We have learnt that the high $T$ expansion corresponds to an expansion in the small entropic field $\phi$. We can expand Eq.~(\ref{QFT3}) at lowest order to find
\begin{align}
\beta{\cal F}_H[\phi] &=-\frac{1}{2} \sum_{vv'}  A_{vv'} \ln \cos \left(\nabla_{vv'}\phi +i\beta H_{vv'}\right) \nonumber \\
 &\simeq \frac{1}{4 }\sum_{vv'} A_{vv'} \left( \nabla_{vv'} \phi + i\beta H_{vv'} \right)^2 \nonumber \\
 &= \frac{1}{2}\sum_{vv'} \phi_v \left(z_v \delta_{vv'}-A_{vv'} \right) \phi_{v'}  \nonumber \\
& - i\beta \sum_{vv'} \phi_v A_{vv'} H_{vv'}  -\frac{\beta^2}{4} \sum_{vv'} A_{vv'} H_{vv'}^2.
\label{QFT4}
\end{align}
The matrix 
\begin{equation}
L_{vv'}=z_v \delta_{vv'}-A_{vv'} 
\label{D2}
\end{equation}
which appears in the third line of Eq.~(\ref{QFT4}) is called in graph theory the {\em Laplacian matrix} of the graph.  It can be written as $\hat L=\hat D-\hat A$ where $D_{vv'}=z_v \delta_{vv'}$ is the  {\em degree matrix}. The Lagrangian matrix is indeed the generalization on a general graph of the discretized Laplacian operator on a lattice. The reader can easily verify that for instance on a square lattice of edge length $a$ when one takes the usual continuum limit for $a\to 0$ one finds $\hat L\to -a^2\nabla^2$. Also, for a generic vector $\zeta_v$ 
\begin{equation}
\mathrm{div}[\nabla \zeta]_v=-\sum_{v'} L_{vv'}\zeta_{v'}
\end{equation}
as one would expect as generalization of the notorious $\vec \nabla \cdot \vec \nabla=\nabla^2$.

We can now write ${\cal F}$ in Eq.~(\ref{QFT2}) at the second order formally as
\begin{align}
\beta{\cal F}_2&=\frac{1}{2} q\hat \Delta q + \frac{1}{2} \phi\hat L \phi - i \phi \cdot \left(q + \beta \mathrm{div}[\hat H]\right)-\frac{\beta^2}{2} |H|^2
\label{QFT5}
\end{align}
where matrices are expressed as operators acting on  $q, \phi$, which are vectors of dimension $N_v$. We have defined 
\begin{equation}
\hat \Delta =(\epsilon \hat 1+\mu \hat V)/T, 
\label{Delta}
\end{equation}
and $|H|^2=\sum_{vv'} A_{vv'} H_{vv'}^2/2$. (From now on we will skip writing the unity operator $\hat 1$: when a scalar is summed to an operator it will be understood that the scalar is multiplied by the unity operator.)

Note that Eq.~(\ref{QFT5}) is correct {\it because} the divergence of $H$ is hereby  defined discretely with respect to vertices by Eq.~(\ref{div}). One might at first imagine that Eq.~(\ref{QFT5}) implies, for instance, that on a square lattice a continuous  solenoidal $H$ would not couple to the spin ice. However, a moment thought shows that  even for an uniform $H$,  its divergence {\it on vertices}, as defined here is necessarily non-zero at the boundaries  of the lattice.

By integrating over $q_v$ we can express the partition function in Eq.~(\ref{QFT}) in the high $T$ approximation as due to a free energy in the entropic field only, or
\begin{align}
\beta{\cal F}_2[\phi]&=\frac{1}{2} \phi \left(  \hat \Delta^{-1} + \hat L \right) \phi - i \phi \cdot \left(\beta \mathrm{div}[\hat H]\right)-\frac{\beta^2}{2} |H|^2.
\end{align}

Using Eqs~(\ref{QFT}), (\ref{QFT5}), we can  find an effective free energy for the charges by integrating instead over the entropic fields $\phi$. To do so we must first consider the spectrum~\cite{brouwer2011spectra} of $L$. 

The following is known from the spectral theory of graphs~\cite{}. $L$ is symmetric and thus has real eigenvalues $\{\gamma(k)^2\}_{k=0\dots k_{\max}}$ with  $k_{\max} \le V-1$, of corresponding $N_v$  eigenvectors $\psi^{\alpha}_v(k)$ (where $\alpha$ counts the eigenvalue degeneracy). 
It is  immediate to verify that  $\gamma^2(0)=0$ for the  uniform eigenvector $\psi_v(0)=1/\sqrt{N_v}$. In a simple and connected graph all other eigenvalues are positive. 

We can go to the new basis, defining $\tilde q^{\alpha}(k)={{\psi^{\alpha}}^*(k)\cdot q}$ and $\tilde \phi^{\alpha}(k)={{\psi^{\alpha}}^*(k)\cdot \phi}$. Then in  Eqs~(\ref{QFT}), (\ref{QFT5}), the integration over $d\tilde \phi(0)$ merely returns a  $\delta(\tilde q(0))$, which in ``real space'' corresponds to $\delta\left(\sum_v q_v\right)$. This ensures  that in the new free energy we sum only over charge configurations of zero net charge--as it should be, since a system of dipoles is charge neutral. All other charge modes have zero net charge. Indeed  for any eigenvector  of $L$ except the one of zero eigenvalue it is true that   $\sum_v \psi_v(k)=0$. This follows immediately from $\sum_v L_{vv'}=z_{v'}-z_{v'}=0$ and $\psi_v(k) =\sum_{v'} L_{vv'} \psi_{v'}(k)/\gamma(k)^{2}$. 

From Gaussian integration of ${\cal F}_2$ in Eq.~(\ref{QFT5}) in the space orthogonal to $\psi(0)$ (where $L$ can be inverted)  we obtain, in absence of field $H$, a new effective free energy for $q$  in the form
\begin{align}
\beta{\cal F}_2[q]&=\frac{1}{2} q\hat \Delta q + \frac{1}{2} q{\hat L}^{-1} q,
\label{QFT6}
\end{align}
which can be interpreted both in real or spectral space. 

The first term contains the energy cost to produce charges and their mutual, {\it physical} interaction. The second term tells  us that $T \hat L^{-1}$ is the kernel of an {\it entropic interaction} between charges. Thus, at quadratic order, the effect of the underlying spin manifold   can be subsumed into a pairwise, entropic interaction that corresponds to $L^{-1}$, which of course is in general the Green matrix of the Laplacian operator (generalization of the Green function of the Laplacian). 

In regular lattices embedded in a linear space (see below),  Eq.~(\ref{QFT6}) implies that in three dimensions (3D) charges interact entropically via a $1/x$ law, as indeed found numerically~\cite{}. In two dimensions (2D) one expects instead a logarithmic interaction. 
This might imply a mismatch in systems of reduced dimensionality, for instance in square ice, whose entropic interaction is the Green function of the 2D laplacian while $V$ is the Green function of the 3D laplacian. We will deal with that mismatch in the next section.

\subsubsection{Charge Correlations}

We can define $\hat W^q$ as the inverse of the kernel of the free energy for the charge from Eq.~(\ref{QFT6}), or
\begin{align}
\hat{W^q}^{-1}=\hat \Delta + \hat {L}^{-1},
\label{W}
\end{align}
and then, from equipartition we have
\begin{align}
 \langle q_v q_{v'}\rangle =W^q_{vv'} 
\label{qcorr5}
\end{align}
For   $\mu=0$ we note that $\hat W^q$ can be written in various ways, including
\begin{align}
\hat W^q =& \hat L(\xi^2\hat L +1)^{-1} \nonumber \\
=&\xi^{-2}- \xi^{-4}\hat G_{\xi},
\label{W2}
\end{align}
where  
\begin{equation}
\xi=\sqrt{\epsilon/T}
\label{xi0}
\end{equation}
is the {\it correlation length at high temperature}, as it was   already appreciated in more specific systems, via other means~\cite{garanin1999classical,henley2005power}.
In the last line we have introduced  the green ``function'' (actually, a matrix) of the screened Poisson equation on a graph, or
\begin{equation}
\hat G_{\xi}= \left(\hat L +\xi^{-2}\right)^{-1}.
\label{G}
\end{equation}

Going to the spectrum of $L$ we find then immediately 
\begin{align}
\langle {{\tilde q}^{\alpha}}{^*}(k) \tilde q^{\alpha'}(k') \rangle&=\delta_{\alpha \alpha'} \delta_{kk'} w(k) \nonumber \\
 &=\delta_{\alpha \alpha'} \delta_{kk'} \frac{\gamma(k)^2}{ \gamma(k)^2\xi^2+1}.
\label{qcorr1}
\end{align}
%
%
In the infinite temperature limit the correlation length becomes zero, and from Eq.~(\ref{qcorr1}) we obtain
\begin{equation}
\langle q_v q_{v'}\rangle \to L_{vv'} ~~\mathrm{for} ~~ T\to \infty,
\label{qcorr2}
\end{equation}
which corresponds to uncorrelated vertices. Let us see why. 
From Eq.~(\ref{qcorr2}) we have  
\begin{equation}
\langle q_v^2\rangle \to z_v~~~\text{for}~~~T\to \infty
\label{qsatur}
\end{equation}
 which  indeed corresponds to the average square charge of uncorrelated vertices, or $ \overline {q^2}_{\text{uncorr}}$ defined as the square charge obtained from counting arguments. Considering vertex multiplicities only, and computing the  average square charge for a vertex of degree $z$ with  each charge $z-2n$  weighted merely by its vertex multiplicity ${z} \choose {n}$, one has indeed 
\begin{equation}
\langle q^2\rangle_{\text{uncorr}}=2^{-z}\sum_{n=0}^z (z-2n)^2 {{z} \choose {n}}=z.
\label{quncorr}
\end{equation}
Note also that Eq.~(\ref{qcorr2}) implies zero correlations among vertices that are not nearest neighbors, but a correlation of $-1$ among neighboring vertices.  Reasonably, even at high $T$, since nearby vertices share a spin, they have in average opposite charges.

We might ask ourselves how is the infinite temperature limit approached. From Eq. (\ref{qcorr1}) we  obtain the series
\begin{align}
\langle q_v q_{v'}\rangle= \left[L \sum_{n=0}^{\infty} \left(-\xi^2 L\right)^{n} \right]_{vv'}, 
\label{qcorr3}
\end{align}
%
which is interesting because $L^n$ can be written as sums of products of $n$ $\hat A$ and $\hat D$ matrices (e.g. $ADDAADAD\dots$) in which each matrix appears at most $n$ times. Considering the definition of $A$, a moment's thought should convince that if $v$, $v'$ are separated by more edges than the number of $A$ matrices in such product, then the $vv'$ element of the product is zero. An obvious notion of  distance between two vertices on a graph is given by the number of edges in the shortest {\it path} (aka graph geodesic) connecting them~\cite{aouchiche2014distance}. It follows that if $v$ and $v'$ are at a distance $d_{vv'}>1$ the first nonzero term of the series in Eq.~(\ref{qcorr3}) is given by
\begin{align}
\langle q_v q_{v'}\rangle= \left(-\xi \right)^{2(d_{vv'}-1)} \left[A^{d_{vv'}}\right]_{vv'} +O\left(\xi^{2d_{vv'}}\right).
\label{qcorr4}
\end{align}
%
%
Interestingly, $A^k_{vv'}$ is known  to be the number of {\it walks} of length $k$ between the two vertices $v$ and $v'$. Thus, the coefficient $ \left[A^{d_{vv'}}\right]_{vv'} $ in Eq.~(\ref{qcorr4}) is the number of walks between the two vertices $v$ and $v'$ of length equal to the distance $d_{vv'}$. It is therefore the number of  geodesics connecting the two vertices. 

By construction our approximation doesn't work for $T\to0$ where the fluctuations of the entropic field diverge.  One could speculate that by including perturbative terms things would not change, except for replacing $\xi(T) \to \xi_r(T)$ where $\xi_r(T)$ is the real correlation length at low $T$. For instance in pyrochlore ice, correlations become screened-algebraic at low $T$, suggesting that a quadratic free energy might work with proper renormalizations of the parameters.   

In the limit $\xi\to\infty$ From Eq.~(\ref{qcorr1}) we find 
\begin{align}
\langle q_v q_{v'}\rangle= \xi^{-2}\delta_{vv'}-\xi^{-4}\left[L^{-1}\right]_{vv'} +O(\xi^6).
\label{trippa}
\end{align}
We immediately understand that such an approach to low $T$ correlations, if it does work, can only work for graphs of even coordination. In such case the equation
\begin{equation}
\xi^2 \sim 1/\langle q^2\rangle~~\text{for}~~ \xi^2 \to \infty,
\label{xir}
\end{equation}
 obtained from Eq~(\ref{trippa}), is not inconsistent a priori. Indeed, in pyrochlore spin ice it corresponds to the experimentally found~\cite{fennell2009magnetic} exponentially divergent behavior of the correlation length in the proximity of the ice-manifold, as we will show in the next section. And because for $v\ne v'$ the correlations tend to $\propto \hat L^{-1}$, we can call them {\it algebraic} in this limit and we say that this limit is {\it critical}. When the graph has vertices of odd coordination (an extensive number of them, in an infinite graph~\cite{}), Eq.~(\ref{xir}) cannot be true.

\subsubsection{Entropic Screening}

Since we have chosen $\mu=0$, all screening is entropic. Such entropic interaction  subsumes the correlations among charges coming from the underlying spin structure. 

One can consider two cases: screening of an external charge and screening of a pinned charge. 

Screening of an external charge would seem prima facie impossible having suppressed any physical charge-charge interaction. In reality, external charges interact with emergent spin ice charges via the coupling of the $H$ field to the spins $S$. To understand it formally, consider the term $\sum_{vv'}S_{vv'}H_{vv'}$ in Eq.~(\ref{Z}). Imagine that we can write a Helmholtz decomposition on the graph, so that $H$ can be represented as $H_{vv'}=\nabla_{vv'}\Psi +H_{vv'}^{\perp}$ where $\Psi_v$ is a field and the second term has no divergence. Then we have
\begin{align}
\frac{1}{2}\sum_{vv'}S_{vv'}H_{vv'}=-\sum_vQ_v\Psi_v+ \frac{1}{2}\sum_{vv'}S_{vv'}H_{vv'}^{\perp},
\label{Hel}
\end{align}
that is, the potential responsible for the divergence-full part of the magnetic field couples to the emergent spin ice charges. 

Let us then call $q_{e}=\mu \mathrm{div}[H]$ the ``external charge" (here $\mu$ is some proper constant with the dimension of an energy, given by the problem) and assume $\sum_v q_{e,v}=0$. Then from Eq.~(\ref{QFT5}), integrating over $\phi$ one has, for $\mu=0$, the {\it screening} of the external charges by internal ones, or
\begin{align}
\langle q\rangle&= -\frac{\mu}{T}\hat W^q  \hat L^{-1} q_e =-\frac{\mu}{\epsilon}  \hat G_{\xi} q_e. 
\label{screen}
\end{align}
Thus, an external charge is screened by the screened green function of the Laplacian with screening length $\xi$. Note that because $\hat L \Psi = - q_e$ we can also write
\begin{align}
\langle q_v \rangle&= -\frac{\mu}{T} \langle q_v q_{v'}\rangle \Psi_{v'}.  
\label{screen}
\end{align}
From Eq.~(\ref{screen}) we obtain the two correlation limits
\begin{align}
\langle q_v\rangle & = - \frac{\mu}{T} \delta_{v  v'} q_{e,v'} +O(\xi^{2}) ~~\mathrm{for}~~ \xi\to 0 \nonumber \\
\langle q_v\rangle & = - \frac{\mu}{\epsilon} \left[L^{-1}\right]_{v  v'} q_{e,v'}+ O(\xi^{-2}) ~~\mathrm{for}~~ \xi \to \infty.
\end{align} 

A pinned charge instead is obtained by summing the partition function only over spin configurations corresponding to a fixed charge, say $q_{\mathrm{pin}}$, on a specific vertex, say $\bar v$. This corresponds to inserting a $\delta(q_{\bar v} -q_{\mathrm{pin}})$ in the functional integral. We will leave the simple calculation to the reader and will report here the result 
\begin{align}
\langle q_v\rangle= \frac{W_{v \bar v}}{W_{\bar v \bar v}} q_{\mathrm{pin}} = \frac{\langle q_v q_{\bar v}\rangle}{\langle  q_{\bar v}^2\rangle}q_{\mathrm{pin}},
\label{screenp}
\end{align}
which does not surprise and correctly yields $\langle q_{\bar v}\rangle=q_{\mathrm{pin}}$. Note also, from Eq.~(\ref{W2}),
\begin{align}
\langle q_v\rangle & = L_{v \bar v} \frac{q_{\mathrm{pin}}}{z_{\bar v}} +O(T^{-2})  ~~\mathrm{for}~~ T \to \infty.
\end{align} 
The difference between the two screenings should not surprise. In the first case an external field interacts locally [see Eq.~(\ref{Hel})] with the spin ice, inducing local effects that then propagate via the charge-charge correlation, while the second case is due to correlations of the free charge with the pinned one.

Note that in the Debye-H\"uckel treatment, the screening length has a form $\xi^2_{\mathrm{DH}}\propto  T/\langle q^2 \rangle$. If we take Eq.~(\ref{xir}) to work, it indeed corresponds to a Debye-H\"uckel screening length for a potential with coupling constant proportional to $T$, which is indeed the case for the entropic potential responsible of the screening. At high temperature, the situation is much different, with $\xi^2=\epsilon/T$. This is because the Debye-H\"uckel approach applies to strong electrolytes that are fully dissociated. In such case, the disorder brought by higher temperature prevents charges from screening, thus increasing the screening radius. In spin ice, instead, charges carry an energy cost  and at higher temperature there are more charges available for the screening.

We have obtained a series of results that are independent of geometry, to highlight the topological nature of the spin ice physics. In the following section we will particularize these notions to specific geometries.

\section{Square (vs. Pyrochlore) Ice} 

The square geometry of spin ice was the first to be realized experimentally, and it was shown that non-ice rule vertices are suppressed  after AC demagnetization~\cite{Wang2006,Nisoli2007,Nisoli2010}. However, because moments impinging in the vertex perpendicularly interact more strongly than moments impinging collinearly, the degeneracy of the ice manifold is lifted and antiferromagnatic (AFM) vertices are energetically favored. This leads to a transition in the Ising class~\cite{Wu1969} to an ordered antiferromagnetic ground state~\cite{Porro2013,zhang2013crystallites,sendetskyi2019continuous} .

Shortly after the first realization, it was proposed~\cite{Moller2006}  that the degeneracy of the ice manifold could be regained by offsetting the height of half of the nanoislands. Ten years after, this proposal was realized~\cite{perrin2016extensive} and is currently under study in real time, real space characterization via photoemission electron microscopy~\cite{farhan2019emergent}. Meanwhile, ice-rule degeneracy has been demonstrated via interaction modifiers placed in the vertices, as nano-dots among islands~\cite{ostman2018interaction}, or nano-holes in a connected spin ice of nanowires~\cite{schanilec2019artificial}. Interestingly, it was recently proposed via micromagnetics that  square realizations made of connected nanowires might privilege ferromagnetic ice rule vertices~\cite{perrin2019quasidegenerate}. 

Below, we will treat many of these proposed scenarios within a unifying framework based on the approach of the previous section. 

\subsection{Square Ice and Its Gauge-Free Duality}

Square ice is a set of classical, binary spins $\vec S_e$ aligned on the $N_l$ edges labeled by $e$ of a square lattice of $N_s=N_l/2$ vertices labeled by $v$. Spins impinging in a vertex form vertex topologies that are usually classified as t-I, \dots, t-IV  and shown in Fig.~2. 

There is in 2D a duality absent in 3D (e.g.\ in pyrochlores). We call it a gauge-free duality because it is related to the rather gravid mathematical fact that in 2D a Helmholtz decomposition has no gauge freedom.

Again, $Q_v[S]$ is the topological charge of the vertex $v$, defined as the number of spins pointing in  the vertex minus those pointing out. In a square lattice, however, we can similarly define  the topological current $I_p[S]$ of the plaquette $p$, as the number of  spins pointing clockwise around the edges of the plaquette minus those pointing counterclockwise. 

 We call $\hat e_1, \hat e_2$ the lattice unit vectors and   $\hat e_3=\hat e_1 \wedge \hat e_2$. Then for each vector $\vec v$  we call  $^{\perp} \vec v=\hat e_3 \wedge \vec v$ the {\em perpendicular} of $\vec v$. Thus for a spin configuration $S$ we can consider its perpendicular $^{\perp \!}S$, by exchanging plaquettes with vertices. For that we can still define  $Q_p[^{\perp} \! \vec S]$ and $I_v[^{\perp}\! \vec S]$. We then  have
\begin{align}
Q_v[\vec S]=-I_v[^{\perp}\! \vec S]~~\text{and}~~ I_p[\vec S]= -Q_p[^{\perp} \! \vec S],
\label{duality}
\end{align}
establishing the equivalence of the ``parallel'' and ''perpendicular'' picture in 2D.

As in the previous section, we call an {\em Ice Manifold}  the subset of the phase space made only of ice-rule~\cite{}  configurations  (or $Q_v[S]=0~\forall v$). It is known that for each such configuration   a unique (up to a constant) {\it height function} $h_{\!\perp }$ can be defined on the plaquettes  by the following 
\begin{equation}
{^{\perp}\!\vec S_e }\cdot \hat{pp'}=h_{\! \perp p'}- h_{\! \perp p},
\label{heighperp}
\end{equation}
 where  $\hat{pp'}$ is the unit vector pointing from plaquette $p$ to $p'$ separated by the edge $e$ (Fig.~1). This is merely a consequence of $^{\perp \!}S$ being current free, or irrotational. 
 
 Similarly, if a spin configuration  has  zero topological current on each plaquette ($I_p[S]=0$ $\forall p$), a unique (up to a constant) height function $h_{|| v}$ can be defined on the vertices  by 
\begin{equation}
\vec S_e \cdot  \hat{vv'} = h_{|| v'}
-h_{|| v}
\label{heighpar}
\end{equation}
where $e$ is the edge connecting the vertices $vv'$.

\begin{figure}[t!]
\includegraphics[width=.5\columnwidth]{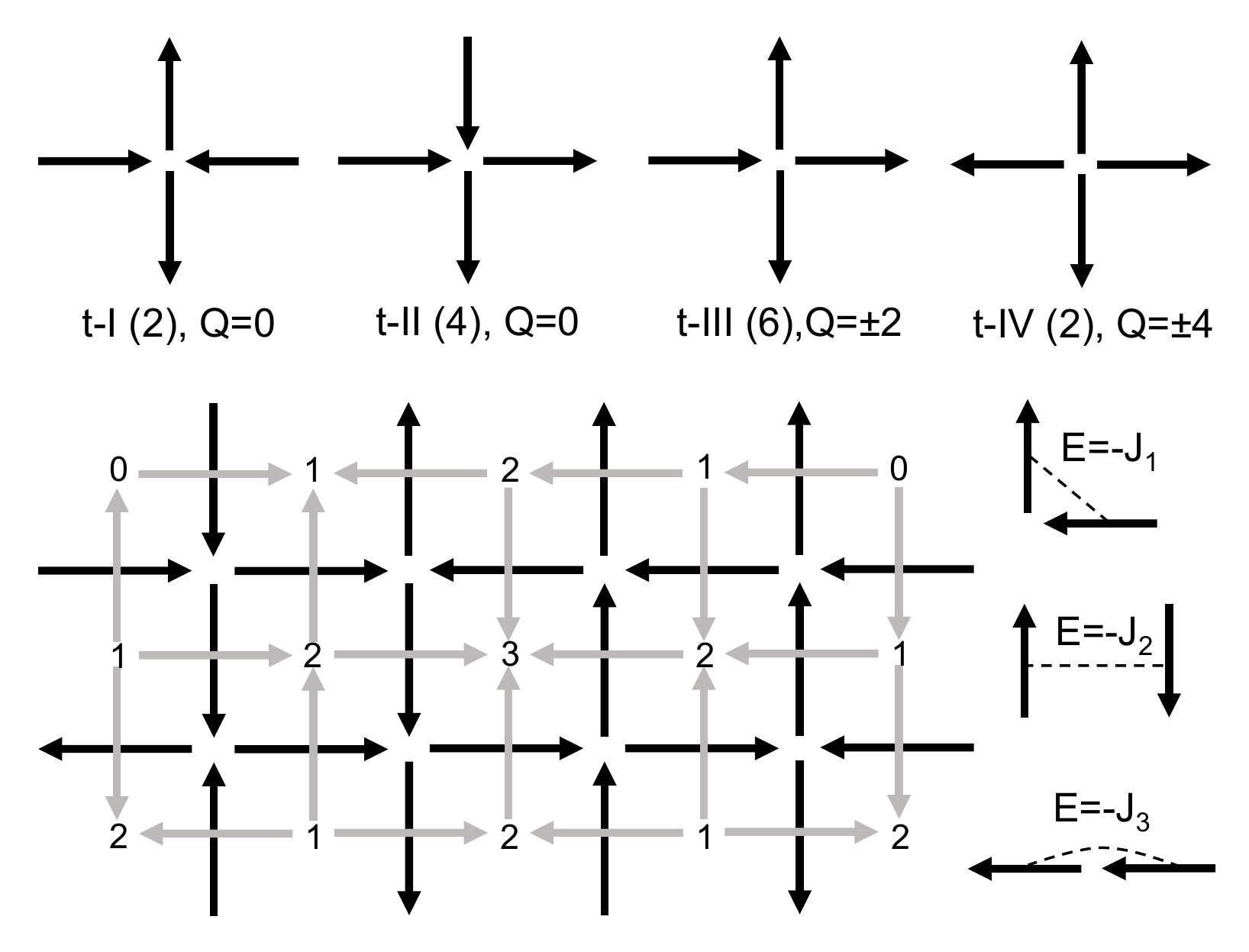}
\caption{Top: the sixteen vertices of square ice can be divided into four topologies, listed with degeneracy in parenthesis and topological charge. Below, an ice-rule obeying configuration of $\vec S$ (black) and its height function $h_{\! \perp}$  built from $^{\perp}\! \vec{S}$ (gray). Also, the coupling constants among spins. }
\label{fig1}
\end{figure}

These concepts are merely the discretized version of the perhaps more familiar continuum analogues
\begin{align}
q[\vec S]& \vcentcolon=  \vec \nabla \cdot \vec S=-\hat e_3 \cdot \vec \nabla \wedge \! {^{\perp}\!{\vec S}}=-j[^{\perp} \!\vec S]  \nonumber  \\
j[\vec S]& \vcentcolon=\hat e_3 \cdot \vec \nabla \wedge \vec S=-q[^{\perp} \! \vec S].
\label{duality2}
\end{align}
to which corresponds the 2D Helmholtz decomposition
\begin{align}
\vec S = \vec \nabla h_{||} - \hat e_3 \wedge \vec \nabla h_{\!\perp}.
\label{Helm}
\end{align}
We call the first (second) term $\vec S_{||}$  ($\vec S_{\!\perp}$). 
From Eqs~(\ref{duality2},\ref{Helm}) $h_{||}$ carries the charge and $h_{\!\perp}$ the current:  
\begin{align}
q &=\Delta h_{||} \nonumber \\
j &=-\Delta h_{\! \perp}.
\label{rhoj}
\end{align}

This formalism realizes  {\em magnetic fragmentation}~\cite{brooks2014magnetic,petit2016observation,canals2016fragmentation} for the square lattice.

The ice manifold is a charge-free state for $S$ and thus is characterized by $\Delta h_{||}=0$ with $h_{\!\perp}$ free to fluctuate, at least within the constrains of its discretized definition. Basically, the disorder is labeled by the choice of currents compatible with zero charge. Instead, $\Delta h_{\!\perp}=0$ defines an ice manifold, or charge-free state, for $^{\perp \!}S$, or equivalently a current-free state for $S$. 

Clearly, the state that has both zero charge and zero currents  (or equivalently, the state that is an ice manifold both for the spin ensemble and its transversal) is ordered, corresponding to all rows and columns of spins aligned in the same direction, and its entropy ($=4$) is finite. However, as we shall see, excited, low energy states above it are reminiscent of the ice manifold topological properties if the energy cost of currents is much smaller than charges.

The ice manifold is clearly a Coulomb phase by our definition above, and thus a topological state of constrained disorder where in lieu of an order parameter the degenerate states are labeled by the field $h_{\! \perp}=0$.  However, in this specific case, more ``Coulombness'' can be proved:  we can see that correlations are Coulomb.

Height models can be described by ``rough'' (degenerate) of ``flat'' (ordered) phases, a jargon derived from the {\em roughening transition} historically associated with these models by various exact mappings~\cite{van1977exactly,chui1976phase}. 

The ice manifold of our  square ice, i.e.\ the six vertex model, is known to be equivalent to a dimer cover model (see ref~\cite{zinn2009six,baxter1982} and references within) and thus in a rough phase~\cite{henley2010coulomb,henley2011classical,henley1997relaxation}. It is a widespread ansatz (widespread, though by no means proved in most cases, and merely adopted via handwaving generalizations see ref~\cite{henley1997relaxation} and references therein) to deal with these systems in the long wavelength limit by ascribing to them a quadratic entropy that in our language reads
\begin{equation}
-{\cal S}=\frac{\zeta}{2}\int \vec \nabla h_{\! \perp} \cdot \vec \nabla h_{\! \perp} d^2x
\label{hen}
\end{equation}
for some constant $\zeta$. (Obviously, the same is true for the current-free manifold that is an ice manifold, or charge-free manifold for $^{\perp} \!S$ by replacing $h_{\! \perp}$ with $h_{\! ||}$).

Note that Eq.~(\ref{hen}) is not obviously unproblematic in a 2D gauge-free theory. In 3D (where the gradient of the height function is replaced by the curl of a vector potential) we would be safe. There,  gauge invariance forbids the proliferation of relevant operators at the fixed point (this is, incidentally, the mathematical reason for having to introduce by hand a Higgs boson in the standard model, that is a new field undergoing symmetry breaking, to generate the mass for the weak force bosons). Eq~(\ref{hen}) merely happens to work in predicting correlations that can also in part be computed exactly~\cite{sutherland1968correlation,baxter1982}.

 Indeed, from Eq.~(\ref{hen}) and equipartition one finds   the correlation function for the heigh function in reciprocal space
\begin{equation}
\langle | h_{\! \perp}(k)|^2\rangle= \frac{\zeta}{k^2}
\label{heighcorr}
\end{equation}
and then from Eq~(\ref{Helm}) the spin correlator as
\begin{equation}
\langle \vec S^*(k) \vec S(k)^2\rangle= \zeta \frac{^{\perp}\! \vec k ^{\perp}\! \vec k}{k^2}=\zeta\left(\hat 1 -\frac{ \vec k \vec k}{k^2}\right)
\label{spincorrheigh}
\end{equation}
 (we have used dyadics) which is purely transversal, and correctly so. Note that  real space correlations of the height function are {\em logarithmic} while  spin correlations are partial derivatives ({\em transversal} derivatives, or $^{\perp}\vec \nabla =\hat e_3 \wedge \vec \nabla $) of the logarithm. As the logarithm is the 2D-Coulomb function, they correspond to the kernel of the dipolar interaction (not in 3D but in 2D, or as it is often said, in 2+1 electromagnetism), or
\begin{equation}
\langle  h_{\! \perp}(x) h_{\! \perp}(y)\rangle= -2\pi \zeta \ln(|x-y|)
\label{heighcorr}
\end{equation}
and
\begin{equation}
\langle  S_{\alpha}(x) S_{\alpha'}(y)\rangle= 2\pi \zeta\left(\frac{\delta_{\alpha,\alpha'}}{|x-y|^2}-2\frac{x_{\alpha}x_{\alpha'}}{|x-y|^4}\right).
\label{heighcorr}
\end{equation}
Compare the above with the case of pyrochlore spin ice where the correlations of the spins is a 3D dipolar interaction.

As temperature is raised charges appears and the charge-full height function becomes $h_{||} \ne 0$, but its coupling to $h_{\perp}$ remains of higher than second order (as we shall see below) and thus many of the topological feature of the ice manifold are retained in the excited state.

\subsection{Energy and States}


Because of this symmetry, absent in a general graph, we will use the following Hamiltonian
\begin{align}
{\cal H}[Q,I]=\frac{\epsilon}{2}\sum_v Q_v^2 +\frac{\kappa}{2}\sum_v I_v^2 + \frac{1}{2}\sum_v Q_vV_{vv'} Q_{v'} 
%
\label{H-n}
\end{align}
that unlike the Hamiltonian of Eq.~(\ref{Hs}) contains a term for currents. It corresponds to  nearest neighbor couplings  (Fig.~1): $J_1=\epsilon-\kappa$ , $J_2=\epsilon$, $J_3=-\kappa$. 
 In magnetic systems we have  $V_{vv'}=\mu/(2\pi |v-v'|)$. We set the lattice constant $a=1$ so that $\mu$ is an energy. In a nanomagnetic realization, assuming a dumbbell model, $\mu/\epsilon\simeq 1-l/a<1$ where $l<a$ is the length of the dumbbell.
 
The Hamiltonian in Eq.~(\ref{H-n}) describes two of the cases discussed in the introduction of this section, and approximates the third. 

If all terms are zero except $\epsilon>0$ then $J_1=J_2$, $J_3=0$ and thus we have a vertex model where all the ice-rule vertices are degenerate. The ground state is an extensively degenerate~\cite{lieb1967residual} ice manifold for $\vec S$. (Equivalently, for the gauge-free duality, if all terms are zero except $\kappa>0$ the ground state is an extensively degenerate ice manifold for $^{\perp}\!\vec S$.).

For  $\epsilon>0$, $\kappa>0$ (and $V=0$) both charges and currents are  suppressed. Because $J_1<J_2$, t-II vertices are promoted over t-I. Because $J_3<0$, t-II vertices want to align and the ground state is ordered. More formally, as the ground state is current-free and charge-free, we have  $\Delta h_{||}=\Delta h_{\! \perp}=0$ which implies a uniform $\vec S$. Thus the ground state is made of t-II vertices ferromagnetically (FM) aligned. Because the ground state is the intersection of topological requirements (charge-free and current-free), there is no phase transition to the FM zero temperature state, just as there is no transition to the ice manifold or the current-free manifold. This FM state has not been investigated. However, as we said, there are cases in which t-II vertices are favored. Ignoring further neighbor interaction, their ground state should be a  line state of subextensive entropy $S= \sqrt{N_l}\ln2$ and might be approximated by our FM case for small $\kappa$ and not too small $T$. 

For  $\epsilon>0$, $\kappa<0$ (and $V=0$) currents are promoted, and the ground state is an ordered antiferromagnetic (AFM) tessellation of t-I vertices which  breaks $Z_2$ symmetry, leading to a second order  transition of the Ising class~\cite{Wu1969,sendetskyi2019continuous} at a certain temperature $T^{\text{afm}}$. This can also be seen by the fact that $J_1>J_2$. Note that this AFM case should also describe spin ices with vertex degeneracy. Indeed these materials are made of dipoles, and vertex-degeneracy does not account for the long-range effect of the dipolar interaction which favors  closed magnetic fluxes and thus promotes topological currents, corresponding to a small negative $\kappa$ in Eq.~(\ref{H-n}).

We  concentrate on the threshold between these cases, corresponding to $\epsilon>0$ and $|\kappa|$ small and we investigate how ice manifold features are retained in this   {\em biased ice manifold}. 

\subsection{Field Theory}
We particularize to the square lattice the treatment of the previous section.
The  partition function from Eq.~(\ref{H-n}) reads
\begin{equation}
Z=\sum_{S} \exp{\! (-\beta {\cal H} +\beta \sum_e \vec S_e\cdot \vec H_e)},
\label{Z-n}
\end{equation}
and from it $\langle \vec S_{e_1} \dots \vec S_{e_n}\rangle=\partial_{\beta \vec H_{e_1} \dots \beta \vec H_{e_n}} \! \! \ln Z$.

To go to continuum fields we insert  in the sum of (\ref{Z-n}) the expression 
\begin{align}
1&=(2\pi)^{-2N_v}\prod_v \int   dq_v d\phi_v\exp\left[i \phi_v\! \! \left(q_v-q_v\right) \right] \nonumber \\
&\times \prod_p \int   di_p d\psi_p\exp\left[i \psi_p\left (j_p-I_p\right) \right]
\end{align}
and then sum over the spins, obtaining
\begin{equation}
Z=\int \left[dq dj\right]  e^{-\beta{\cal H}[q,j]} \tilde \Omega[q,j]
\label{Z2-n}
\end{equation}
where $\left[dq dj\right]=(2\pi)^{-N_v} \prod dq_v \prod d j_p$. 
The density of states for  $q_v, j_p$ is  given by
\begin{equation}
\tilde \Omega[q,j]=\int [d\phi d\psi] \Omega[\phi, \psi]e^{i \sum_v  q_v \phi_v+i\sum_p  j_p \psi_p},
\label{Omegatilde-n}
\end{equation}
which is the functional Fourier transform of 
\begin{equation}
\Omega[\phi, \psi]=2^{N_l}{\prod_{{\langle vv' \rangle}}}\cos \left(\nabla_{vv'}\phi +\nabla_{pp'}\psi- i \beta H_{vv' } \right)
\label{Omega-n}
\end{equation}
%
where the product runs on all the edges $e={\langle vv' \rangle}$ once, and as before $\nabla_{vv'}\phi \vcentcolon=  \phi_{v'}- \phi_{v}$,  $\nabla_{pp'}\psi \vcentcolon= \psi_{p'}- \psi_{p}$, $ H_{vv' } \vcentcolon=  \vec H_e\cdot \hat{vv'}$, while $\hat{vv'}=\hat e_z \wedge \hat{pp'}$. 
Note that by construction $\langle Q_{v_1} \dots Q_{v_n}\rangle=\langle q_{v_1}\dots  q_{v_n}\rangle$, and the same holds for $I$ and $j$.

We have again gone from discrete variables to a theory of continuous emergent topological charges and currents constrained by an entropy 
\begin{equation}
S[q,j]=-T\ln \tilde \Omega[q,j]
\end{equation}
 which conveys the effect of the underlying spin structure. Equivalently, in the language of field theory,  charges  $q_v$ and currents $j_p$ interact  {\em entropically} via the ``bosonic fields'' $i \phi_v$, $i \psi_p$ of  free energy  
 \begin{equation}
 {\cal F}[\phi, \psi]=-T\ln \Omega[\phi, \psi].
 \end{equation}

\subsubsection{High $T$ Behavior}


Because we are interested  in the monopole liquid below the ice manifold threshold $T\simeq 2 \epsilon$ but above other possible transitions, 
we can perform an high $T$ approximation which, like in the previous section, corresponds to the lowest order in the entropic fields $\phi, \psi$. 


We  expand $\ln \Omega[\phi, \psi]$ at quadratic order and Fourier transform via 
\begin{equation}
g_x=\int_{BZ}  \tilde g(\vec k)e^{-i \vec k \cdot x }\frac{d^2k}{(2\pi)^2},
\end{equation}
 where BZ is the Brillouin Zone, 
 $g$ is a generic field, and $x=v,p,l$  represents vertices, edges or plaquettes. We obtain 
\begin{align}
Z_2=\int [dq dj][ d\phi d\psi]\exp\! \! \left(-\int_{\text{BZ}} \beta {\cal F}_2(\vec k) \frac{d^2k}{(2\pi)^2} \right) \nonumber \\
\label{Z2-n}
\end{align}
with
\begin{align}
&{\cal F}_2[\tilde q, \tilde j, \tilde \phi, \tilde \psi]  = \frac{\epsilon+\mu \tilde V}{2} \left|\tilde{q}\right|^2 + \frac{\kappa}{2} |\tilde{j}|^2 + \frac{T}{2}  \gamma^2 \! \! \left(| \tilde   \phi |^2 + | \tilde \psi |^2 \right)  \nonumber \\
& - iT \! \left(\tilde{q}^*\tilde{\phi} + \tilde{j}^*\tilde{\psi}\right) + i  \!  \left(^{\perp}\!\vec{\gamma} \tilde{\psi}^*  -   \vec \gamma \tilde{\phi}^* \right)\cdot \vec{\tilde{H}}   - \frac{\beta}{2}  \left| \vec{\tilde{H}}\right|^2.
\label{f2-n}
\end{align}
We have introduced the vector  
\begin{equation}
\gamma_{\alpha} = 2\sin(k_{\alpha}/2)
\end{equation}
for $\alpha=x,y$ and $^{\perp} \! \vec{\gamma}=\hat e_3 \wedge \vec \gamma$. In the long wavelength limit: $\vec \gamma \simeq \vec k+ O(k^3)$. 

Note that Eq.~(\ref{f2-n}) is both a particularization of Eq.~(\ref{QFT5}) to a square lattice and its generalization to include currents. Clearly, in the language of the previous section, $\gamma(\vec k)^2$ is the spectrum of the square lattice as a graph, while  $-i\vec\gamma \cdot \vec{\tilde w}$, $-i\vec\gamma \wedge \vec{ \tilde w}$ are the generalized divergence and curl respectively, in  moment space, for a generic field $\vec w$ of the graph. $\tilde V(k)$ is the Fourier transform of $V_v$ and $\tilde V(k)\sim 1/q$ as $q\to 0$. 

We see that the entropic fields for currents (resp. charges) couples to the curl (resp. divergence) of the external field. Importantly, no $\phi  \psi$ cross term survives, ensuring spin fragmentation: charge and currents are independent at second order.  

%

Integrating $Z_2$ over $\tilde \phi, \tilde \psi$ for $\tilde H=0$ returns the effective free energy for $q$ and $j$ in absence of external field
\begin{align}
&{\cal F}_2 [q, j] = {\cal F}_2 [q] +{\cal F}_2 [j]
\label{f2rho}
\end{align}
with
\begin{align}
&{\cal F}_2 [q] = \frac{1}{2}\! \! \left( \epsilon + \mu \tilde V  +\frac{T}{\gamma^2} \right) |\tilde{q}|^2  \nonumber \\
&{\cal F}_2 [j] = \frac{1}{2}\! \! \left( \kappa +\frac{T}{\gamma^2} \right) |{\tilde j}|^2.
\label{f2rho}
\end{align}
%

We  learn from Eqs~(\ref{f2rho}) that at large distances ($\gamma^2\simeq k^2$) the entropic charge-charge  interaction is logarithmic, or 
\begin{equation}
{V_e(v-v')\simeq -2\pi q_v q_{v'}T\ln|v-v'|},
\label{Ve}
\end{equation}
and the same is true for currents. In three dimensions, the entropic interaction would instead be $\simeq1/x$, and it would merely alter the coupling constant $\mu$ for the  monopole-monopole interaction by adding a positive temperature dependent term, as already found numerically~\cite{castelnovo2011debye,chern2014realizing}. None of this is especially surprising given that both are the kernel of the inverse operator of the Laplacian in 2D and 3D respectively, often called 2D-Coulomb and 3D-Coulomb functions. However, the {\em real}, that is not merely entropic, interaction among charges in 2D remains a 3D-Coulomb. This mismatch has significant consequences, as we will see.  
 
In a quadratic theory the correlation functions are the inverse of the kernel of the free energy. From Eq.~(\ref{f2rho}) we obtain them as 
\begin{align}
&\langle | \tilde  q(\vec {k}) |^2\rangle = \gamma(\vec k)^2 \tilde \chi_{||}(k) \nonumber \\
&\langle | \tilde  j(\vec {k}) |^2\rangle = \gamma(\vec k)^2 \tilde \chi_{\!\perp} (k)
\label{rhocorr}
\end{align}
where $\tilde \chi_{||}(k), \tilde \chi_{\! \perp} (k)$  are given by
\begin{align}
\tilde \chi_{||}(k)^{-1}&={1 + {\xi_{||}}^2 \gamma(\vec k)^2\left[1+ \bar q\tilde V(k) \right]} \nonumber \\
\tilde \chi_{\!\perp}(k)^{-1}&={1 +{ \xi_{\! \perp}}^2 \gamma(\vec k)^2}, 
\label{taus}
\end{align}
and $\bar q=\mu/\epsilon$, $\xi_{||}=\sqrt{\epsilon/T}$, $ \xi_{\! \perp}=\sqrt{\kappa/T}$. 
We will show later that $\xi_{\! \perp}$ is the high $T$ correlation length for the currents while  $\xi_{||}$ is the correlation length for the charge when $\mu=0$.  For $\mu\ne 0$ things become considerably more complex. 

Note that $\tilde \chi_{||}(k), \tilde \chi_{\! \perp} (k)$ are the the longitudinal and perpendicular susceptibilities (multiplied by $T$) for $\tilde S_{||}$, $\tilde S_{\perp}$. That can be seen immediately by performing the Gaussian integral in Eq.~(\ref{Z2-n}). Remembering that $ \hat \gamma_{\alpha} \hat \gamma_{\alpha'} + {^{\perp} \! \hat \gamma_{\alpha}} \! \!{^{\perp} \! \hat \gamma_{\alpha'}} = \delta_{\alpha,\alpha'}$  one obtains 
\begin{align}
{\cal F}_2 = \frac{1}{2} (\beta \vec{\tilde H}^*)\cdot \left( {\hat \gamma \hat \gamma}{ \tilde \chi_{||}} +  { {^{\perp} \! \hat \gamma} {^{\perp} \! \hat \gamma}}{\tilde \chi_{\! \perp}}\right)\cdot (\beta \vec{\tilde H})
\label{ftot}
\end{align}
where $\hat \gamma$ is  the unit vector of $\vec \gamma$ and we have used dyadics in $\hat \gamma$. The first term in Eq.~(\ref{ftot}) relates the magnetization to the divergence of the external field $H$ while the second to its curl, demonstrating that $\tilde \chi_{||}(k), \tilde \chi_{\! \perp} (k)$ are indeed the longitudinal and perpendicular susceptibilities.   From Eq.~(\ref{ftot}) one finds immediately
the spin correlations:
\begin{equation}
\langle \tilde S^*_{\alpha}(\vec k) \tilde S_{\alpha'}(\vec k)\rangle=  \hat \gamma_{\alpha} \hat \gamma_{\alpha'} \tilde \chi_{||}+   {^{\perp} \! \hat \gamma_{\alpha}} \! \!{^{\perp} \! \hat \gamma_{\alpha'}}\tilde \chi_{\! \perp}.
\label{spincorr}
\end{equation}

In the limit $T\to \infty$, Eq.~(\ref{spincorr}) returns  $\langle \tilde S_{\alpha}(\vec k) \tilde S_{\alpha'}(\vec k)\rangle \to \delta_{\alpha, \alpha'}$   corresponding to uncorrelated spins. Finally, from the spin correlations we can obtain the magnetic structure factor
\begin{equation}
\Sigma_m(\vec k)=^{\perp}\! \!\vec{k}\cdot \langle \vec{\tilde{S}}(\vec k) \vec{\tilde{S}}(\vec k) \rangle \cdot ^{\perp}\!\!\vec{k}
\label{sigma}
\end{equation}
which will come in use later.

Note also that we started with Eq.~(\ref{charge}) that defines the charge as a divergence of the spins. Similarly the currents are the curl of the spins. And indeed we have from Eq.~(\ref{spincorr}) and (\ref{rhocorr})
\begin{align}
&\langle | \tilde  q(\vec {k}) |^2\rangle = \vec{\gamma}\cdot \langle \vec{\tilde{S}}(\vec k) \vec{\tilde{S}}(\vec k) \rangle \cdot \vec{\gamma} \nonumber \\
&\langle | \tilde  j(\vec {k}) |^2\rangle = ^{\perp}\! \!\vec{\gamma}\cdot \langle \vec{\tilde{S}}(\vec k) \vec{\tilde{S}}(\vec k) \rangle \cdot ^{\perp}\!\!\vec{\gamma}
\label{rhocorrS}
\end{align}
%


Equation~(\ref{spincorr}) implies at quadratic order the following effective free energy for the spins 
\begin{equation}
\beta {\cal F}_2[S]=\int_{BZ} \! \left(\frac{1}{2 \tilde \chi_{||}} {\tilde S}_{||}^*\cdot {\tilde S}_{||} + \frac{1}{2 \tilde \chi_{\!\perp}} {\tilde S}_{\!\perp}^* \cdot {\tilde S}_{\!\perp}\right) \frac{d^2k}{(2\pi)^2} 
\label{FS}
\end{equation}
where ${\tilde S}_{||}=\vec{\tilde S} \cdot \hat \gamma$ is the charge-full, current-fee part of the magnetization whereas ${\tilde S}_{\! \perp}=\vec{\tilde S} \cdot \hat \gamma_{\! \perp}$ is  charge-free, current-full. 

\subsubsection{Low $T$ Behavior}

We have deduced our equations in the limit of high $T$. Can we use them to say something at low $T$?  We can modify our high $T$ formalism  to reach low $T$, at least for $\kappa \ge 0$ where there are no transitions at $T>0$. 

At low $T$ fluctuations of the entropic fields diverge, suggesting a proper study of the higher orders of $\Omega(\phi,\psi)$. 

We can however take a shortcut and postulate that corrections do not change the functional form of ${\cal F}_2$ in Eq.~(\ref{f2rho}) but merely renormalize its constants by introducing a temperature dependence. 
Proceeding heuristically, we note that 
\begin{equation}
\langle q^2\rangle=\int_{\text{BZ}} \langle | \tilde  q(\vec {k}) |^2\rangle \frac{d^2k}{(2\pi)^2},
\label{q2}
\end{equation}
 and from Eq.~(\ref{rhocorr}) we obtain $\langle q^2\rangle \to 4$  for $T \to \infty$, which is correct. Indeed $\langle q^2\rangle = 4$  is the value deducible from a multiplicity argument ($2^2/2+4^2/8=4$). We have in fact already demonstrated this more generally for any graph in Eq.~(\ref{qsatur}). 
 
 If we still assume $\xi_{||}^{2} \to \infty $ as $T \to 0$, then Eq.~(\ref{q2})  implies
\begin{equation}
\langle q^2\rangle \sim  \xi_{||}^{-2}~~~ \text{for} ~T\to 0,
\label{DH}
\end{equation}
 which solves our problem at least at a practical level.

Indeed $\langle q^2\rangle$ is extremely well  approximated by the naive $\overline{q^2}\sim (16/3)\exp(-2\epsilon/T)$  
 %
 %
computed by assuming uncorrelated vertices. We have thus
 \begin{equation} 
 \xi_{||}\sim \frac{\sqrt{3}}{4}~\! \exp\left({\frac{\epsilon}{T}}\right) ~~~ \text{for}~T\to 0.
 \end{equation}
This exponential behavior of the correlation length when the system approaches the ice manifold was indeed suggested experimentally by analyzing the pinch points in the structure factor of pryrochlore ice~\cite{fennell2009magnetic}. It points to the topological nature of the $T=0$ manifold described at the beginning.  

We can therefore keep our quadratic free energy at low $T$ as long as $\epsilon$ is renormalized to $\epsilon \to \epsilon_r(T) \sim T/\langle q^2\rangle$, which is of rather intuitive meaning. Indeed, a look at Eq.~(\ref{Omega-n}) shows that $\Omega[\phi,\psi]$ is periodic in the gradient of the entropic fields. That is due to is origin from delta functions summed over the spins, and which have two roles: one is to convey the entropic interaction, and the other is to preserve the information that charges are discrete.

We can show that in a simple way and in all generality in coordination $z$. Considering the bipartite lattice of vertices $A-B$. Then when $\kappa=0$ (and thus integration over currents enforces $\psi=0$)  and $\vec H=0$,  Eq.~(\ref{Omega-n})  becomes
\begin{equation}
\Omega[\phi] =2^{N_l}\prod_{v_a}\omega[\phi_{v_b}]
\label{Omega-n2}
\end{equation}
with
\begin{equation}
\omega[\phi_{v_a}] =\prod_{v_b \in \partial v_a} \cos(\phi_{v_a}- \phi_{v_b})\simeq \cos(\phi_{v_a}-\overline \phi_{b})^z
\label{Omega-n3}
\end{equation}
where $\partial v_{A}$ is the set of vertices connected to $v_{A}$ and we have performed a mean field approximation $\phi_{v_B}\simeq \overline \phi_B$ assuming that the entropic field on vertices $B$ surrounding $v_{A}$ is about the same. Then one has immediately
\begin{equation}
\int d\phi_{v}e^{-i \phi_{v} q_{v}} \omega[\phi_{v} ]=e^{-i \overline \phi_{b}q_{v} } \tilde \omega[q_{v}]
\end{equation}
where
\begin{equation}
 \tilde\omega [q]=2^{-z}\sum_{n=0}^{z} {{z} \choose {n}} \delta(q-q_n)
\end{equation}
merely restricts  $q$ to the only possible charges $q_n=(2n-z)$ each with proper multiplicity ${{z} \choose {n}}$.

Note that in our high temperature limit we have taken $\nabla \phi$ to be small. We thus neglected the periodicity needed to constrain the magnitude of charges. In this scenario, the ``dressing''    $\epsilon \to \epsilon_r(T) \sim T/ \overline{q^2}$ takes care of that constraint at low $T$ (that is, already at the level of equipartition), while $\phi$ purely transmits entropic correlations coming from the underlying spin structure.


In fact, we can  reconcile our equations with the phenomenological entropy commonly used in height models at zero $T$, and illustrated at the beginning, in Eq.~(\ref{hen}). By integration over charges and currents in Eq.~(\ref{Z2-n}) one finds the free energies for the entropic field 
\begin{align}
\beta{\cal F}_2 [\phi] = \frac{1}{2}\! \! \left( \gamma^2 + \xi_{||}^{-2}\right) |\phi|^2 
%
\label{f2phi}
\end{align}
from which comes the correlations
\begin{align}
\langle |\phi(k)|^2\rangle=\frac{1}{\gamma(k)^2 +\xi_{||}^{-2}}
\label{phicorr}
\end{align}
This corresponds in real space and long wavelength limit to
\begin{align}
\beta{\cal F}_2 [\phi] \simeq \frac{1}{2}\!  \left( \vec \nabla \phi \cdot \vec \nabla \phi + \xi_{||}^{-2} \phi^2 \right).
\label{f2phire}
\end{align}
When $\xi_{||}^{-2}=0$, we regain therefore the standard, algebraic height function entropy of Eq~(\ref{hen}) with the assumption
\begin{equation}
\phi \to {\sqrt{\zeta}} ~h_{\perp}~~\text{for}~~T\to0.
\label{RG0}
\end{equation}
One should make no mistakes here. The identification $\phi = h_{\perp}$ is certainly true in the pure ice manifold. One would be tempted by the symmetry of the formalism to extend it to any temperature along perhaps with $\psi=h_{||}$. And  one would thus be mistaken. 

Indeed, consider $\kappa=0$. Then, integration over the currents $j$ in Eq~(\ref{f2-n}) implies $\psi=0$. However, for $T>0$  charges exist, and thus $h_{||} \ne 0$. So, at non-zero $T$ clearly $\phi \ne h_{\perp}$. Rather $\phi$ must contain informations from both height functions. Thus, a more general relationship must exist in the form of 
\begin{align}
h_{\perp}& =\nu_{\perp}(\phi,\psi) \nonumber \\
h_{||}& =\nu_{||}(\phi,\psi).
\label{RG}
\end{align}
where the $\nu$s depend on $\kappa, \epsilon, T$. 

To address these issues properly, a renormalization group approach to our formalism would be needed, but it would exceed the more practical scope of this article. We will report on that in future work, as well as  a more elegant way to frame Eqs~(\ref{RG}), (\ref{Helm}) in terms of  Wirtinger derivatives of holomorphic functions.

From here on we will merely enforce Eq~(\ref{DH}) and for simplicity use $\zeta=1$ (i.e\ we do not dress the entropic fields) when approaching the low-temperature behavior. Note that when $\kappa>0$ the same can be done for $\xi_{\! \perp}\sim 1/ \overline {j^2}$. We will let the reader verify that all previous considerations for $\epsilon$ apply for $\kappa$ and indeed $\overline {j^2}$ can be computed exactly as $\overline {q^2}$ by merely exchanging $\epsilon$ with $\kappa$. That is all rather obvious given the gauge-free duality.

Finally, in all our plots we will practically interpolate the behavior of $\xi_{||}$ such that it is $\xi_{||}^2\sim {\epsilon/T}$ for $T \to \infty$ and $\xi_{||}^2\sim 1/ {\overline{q^2}}$ for $T \to 0$ via the empirical formula
\begin{equation}
\xi_{||}^{2 C}=\left(\frac{1}{\overline{q^2}}-\frac{1}{4}\right)^C+\left(\frac{\epsilon}{\epsilon+T}\right)^C
\label{empiric}
\end{equation}
which, when used into Eq.~(\ref{q2}) returns a very good agreement with results of Monte Carlo simulations for square ice (which we performed but do not reported here) for $C=10$.

We have now collected enough tools to analyze the various states of our Hamiltonian.

\subsection{Spin Ice ($\epsilon>0$, $\kappa=0$)}

\subsubsection{Pure Degenerate Spin Ice ($\epsilon>0$, $\mu=\kappa=0$)}

The ground state is the degenerate six-vertex model~\cite{lieb1967residual} and there is no interaction among the charges ($\mu=0$). This, as we shall see, leads to entropic exponential screening of charges at any $T>0$ and algebraic screening and correlations at exactly $T=0$. $T=0$ is therefore a critical point for a topological transition to a topologically ordered state. 

From Eq.~(\ref{taus}) we see that for $\kappa=0$ $\chi_{\perp}=1$ and from Eq.~(\ref{rhocorr}) we get that currents are completely uncorrelated, even though charges are. (We have already mentioned that while this might be the case in our vertex model, it is also not very physical in a system of dipoles where the magnetic fluxes want to close, and we will explore the case $\kappa<0$ later.)

Equation~(\ref{rhocorr}) for $\mu=0$ can be rewritten as
\begin{equation}
\langle |\tilde q(k)|^2 \rangle=\xi_{||}^{-2}\left(1-\frac{1}{1+\xi_{||}^2 \gamma^2}\right)
\label{pappa}
\end{equation}
where the first term Fourier-transforms to a Kronecker delta, whereas the second
 implies at large distance the charge correlation
\begin{align}
&\langle q_v{_1} q_{v_2}\rangle = -\frac{1}{2\pi \xi_{||}^{4}} K_0\left({|v_1-v_2|}/{\xi_{||}}\right). 
\label{K0charge}
\end{align} 
Because the modified Bessel function is exponentially screened, or $K_0(x)\sim \sqrt{\pi/2x}\exp(-x)$, $\xi_{||}=\sqrt{\epsilon/T}$ is a bona fide correlation length at large $T$, as already appreciated via other means~\cite{garanin1999classical}. 

The situation is the same in pyrochlore in reciprocal space, but in real space $K_0$, which is the screened 2D-Coulomb, is replaced by a screened 3D-Coulomb  or  $\exp(-|v|/\xi_{||})/|v|$. This Coulomb behavior is therefore rather general, and indeed we found it to be true in the previous section for a general graph, in Eq.~(\ref{G}), using the screened Green operator of the graph Laplacian.

Correlations are then related to charge screening. In artificial realizations it is possible to pin a charge $Q_{\mathrm{pin}}$ in $v_0$.  Then the pinned charge generates a charge distribution 
\begin{align}
\langle q_v\rangle=\frac{\langle q_v q_{v_0}\rangle}{\langle  q^2\rangle}q_{\mathrm{pin}}
\label{screenp}
\end{align}
and for $\mu=0$ the screening comes entirely from the entropic interaction, which is the kernel of the inverse Laplacian operator.  

Note that at low $T$ $ \xi_{||}$ from Eq.~(\ref{DH}) is formally the expected Debye-H\"uckel  length for a potential whose coupling constant is $\propto T$, as it is indeed the case for our entropic potential. Instead, at large $T$  monopoles can be created easily for a tighter screening, explaining $\xi_{||}^2=\epsilon/T \to 0$ for $T\to \infty$, something that naturally escapes the Debye-H\"uckel approach to strong electrolytes. 

Physically, an exponential screening/correlation of charges is related to the absence of dipolar configurations. Instead algebraic correlations can be considered a failure of screening due to charges that prefer to remain bound. More on that in the following subsection. 

\begin{figure}[t!]
\includegraphics[width=.5\columnwidth]{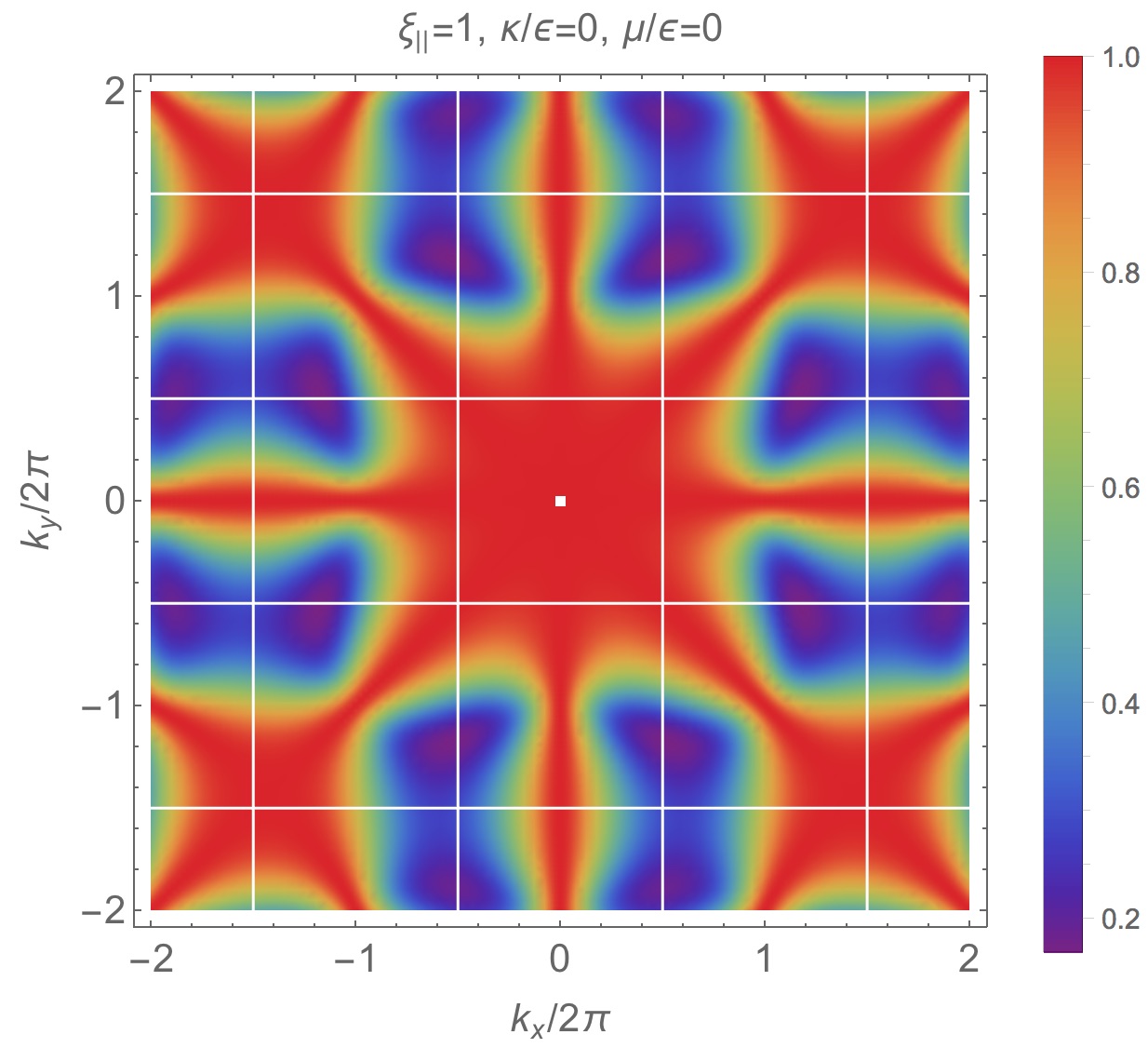}
\includegraphics[width=.5\columnwidth]{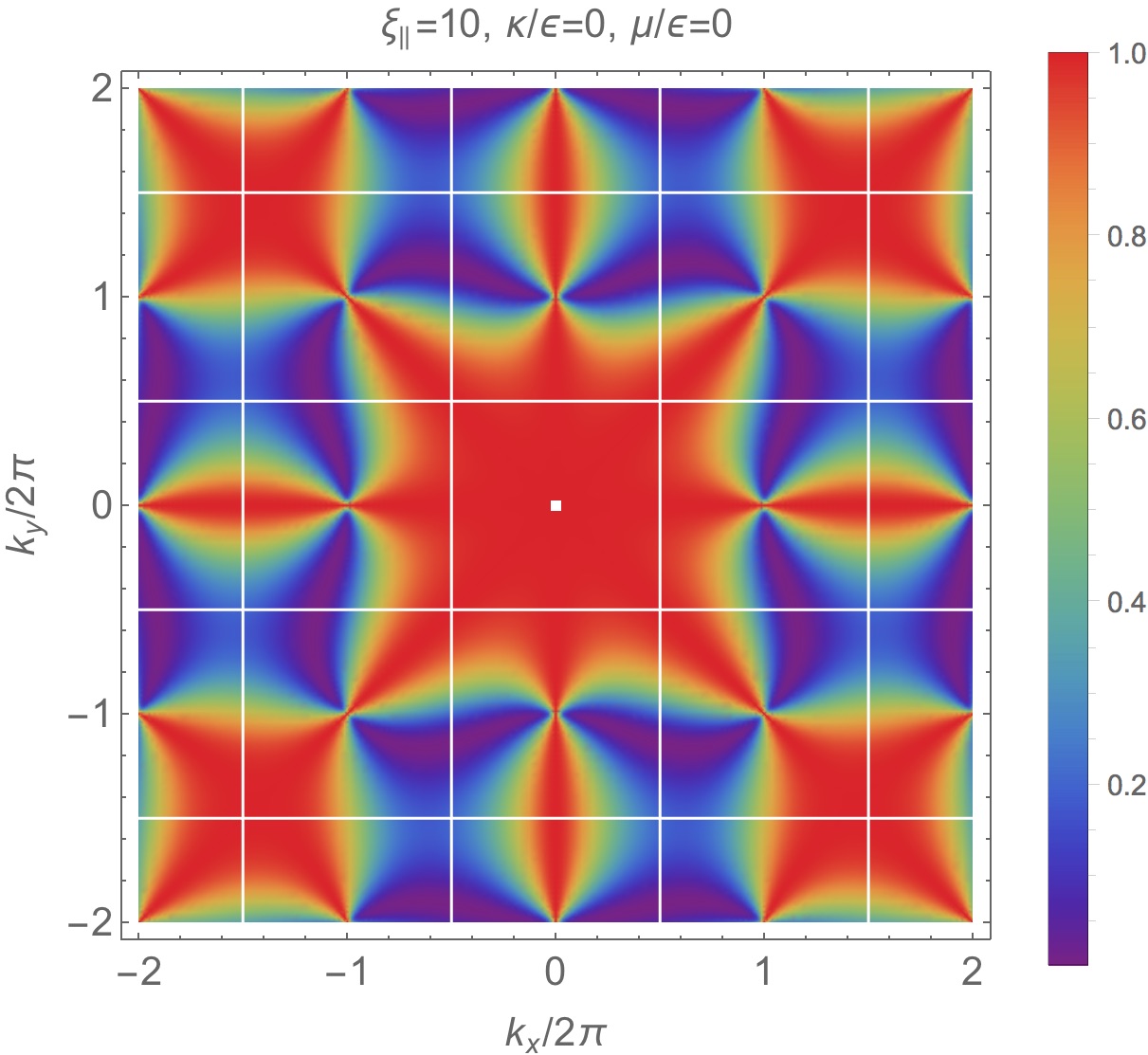}
\includegraphics[width=.5\columnwidth]{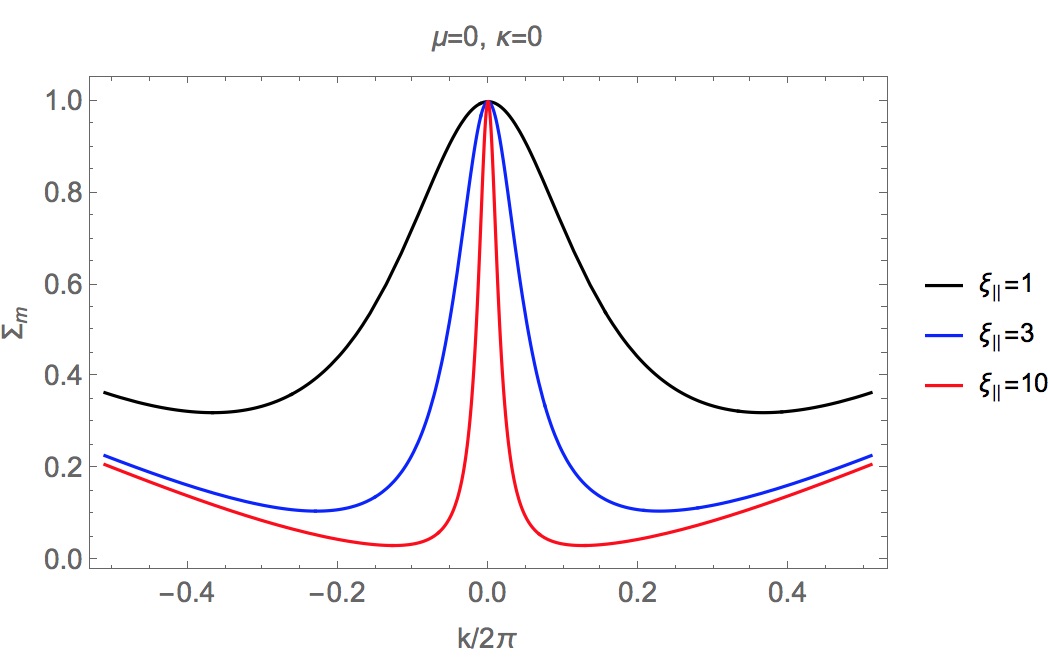}

\caption{Structure factors at high (top) and low (middle) temperature plotted from Eq.~(\ref{sigma}), for degenerate  square ice without monopole interaction. Bottom: structure factor across a pinch point.}
\label{IMsf}
\end{figure}

Considering the spin correlations, Eq.~(\ref{spincorr}) particularizes now to
\begin{align}
\langle \tilde S^*_{\alpha}(\vec k) \tilde S_{\alpha'}(\vec k)\rangle &= \delta_{\alpha,\alpha'} -  \hat \gamma_{\alpha} \hat \gamma_{\alpha'} \left(1- \tilde \chi_{||}\right) \nonumber \\
&=\delta_{\alpha,\alpha'} - \frac{ \gamma_{\alpha} \gamma_{\alpha'}}{\xi_{ ||}^{-2} + \gamma^2}
\label{spincorrIM}
\end{align}
 and in real space  at  distances larger than lattice discretization we have
\begin{equation}
\langle S_{\alpha}(\vec x)  S_{\alpha'}(\vec 0)\rangle\simeq 2\pi \partial^2_{\alpha \alpha'} K_0(|\vec x|/\xi_{||}),
\end{equation}
which of course is {\em not} algebraic, as expected at non-zero temperature. It is only algebraic for $T=0$ when  for the pure ice manifold we obtain
\begin{align}
\langle \tilde S^*_{\alpha}(\vec k) \tilde S_{\alpha'}(\vec k)\rangle_{\text{IM}}=  {^{\perp} \! \hat \gamma_{\alpha}} \! \!{^{\perp} \! \hat \gamma_{\alpha'}}=\delta_{\alpha,\alpha'} - \frac{\vec \gamma_{\alpha} \vec \gamma_{\alpha'}}{\gamma^2}
\label{spincorrIM}
\end{align}
an equation already well known in pyrochlore spin ice~\cite{henley2005power,isakov2004dipolar,youngblood1981polarization,huse2003coulomb} and obtainable merely from the height function formalism, as we have already shown.

At long wavelength, the spin correlations in real space in the pure ice manifold are thus the  kernel of a dipolar interaction in 2D, or 
\begin{equation}
\langle S_{\alpha}(\vec x)  S_{\alpha'}(0)\rangle_{\text{IM}}\simeq 2\pi \partial^2_{\alpha \alpha'} \ln(|\vec x|),
\end{equation}
obtained from as partial derivatives of the 2D Coulomb interaction, as we had seen already at the beginning.

The correlations in Eq.~(\ref{spincorrIM}) are completely transversal and lead to the well known pinch points in the structure factor $\Sigma_m(\vec k)$, which we plot in Fig.~\ref{IMsf}. 

We have so far been concerned with long wavelength behavior. We end this subsection by noting that we can gain some knowledge of  screening at small distances by approximating around the $K=(\pm 1,\pm 1)\pi$ points of the BZ. There, $\gamma(k)^2$ is maximum and $\gamma(\vec k+K)^2=8-k^2$. This leads to a screening function
\begin{equation}
\langle |\tilde q(K+k)|^2\rangle\simeq\xi_{||}^{-2}\left(1-\frac{\xi_{||}^{-2}}{\xi_p^{-2}- k^2}\right)
\end{equation}
where $\xi_p^2 = \xi_{||}^2/(1+8 \xi_{||}^2)$. This suggests that for small $v$ the correlation (or equivalently screening) has a sign alternation with the Manhattan distance on the graph and an envelope function $E(\xi_p |v|)$ of periodicity $\xi_p/2\pi$, or of the form
\begin{align}
\langle q_v{} q_{0}\rangle = C +(-1)^{v_x+v_y}E(\xi_p |v|)
\end{align}
where C is a constant. We should note, however, that monopoles are characterized not only by a charge, but also by a net magnetic moment, which was not taken into account here. Thus, at short distances the screening would be anisotropic, just as monopoles are. In a future work we will incorporate that degree of freedom to study at short length or equivalently around the $K$ points of the BZ.

\begin{figure}[t!]
\includegraphics[width=.5\columnwidth]{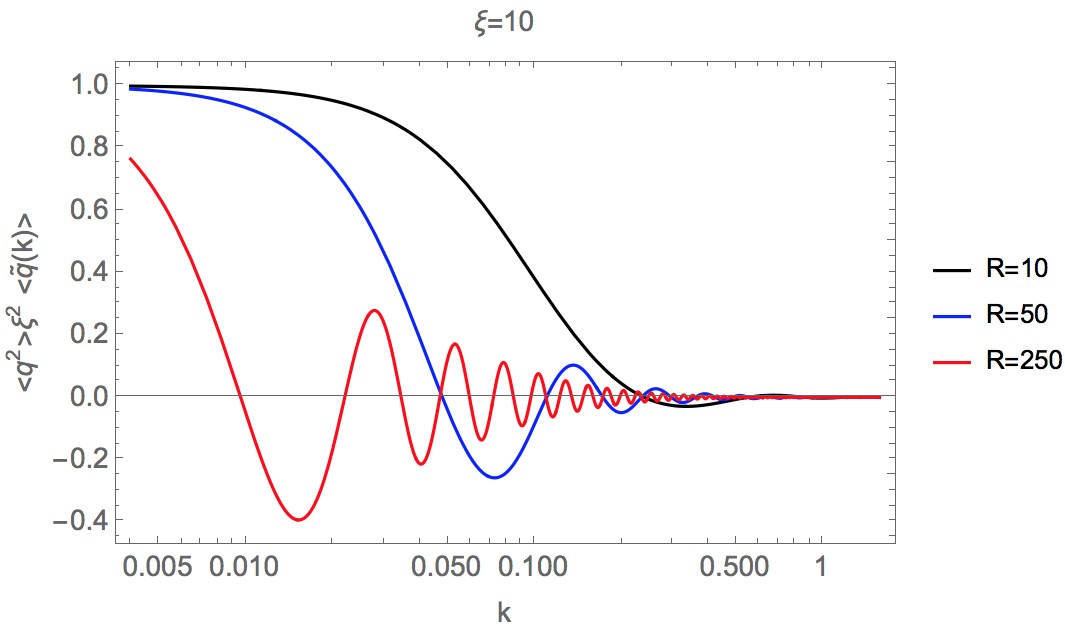}
\includegraphics[width=.5\columnwidth]{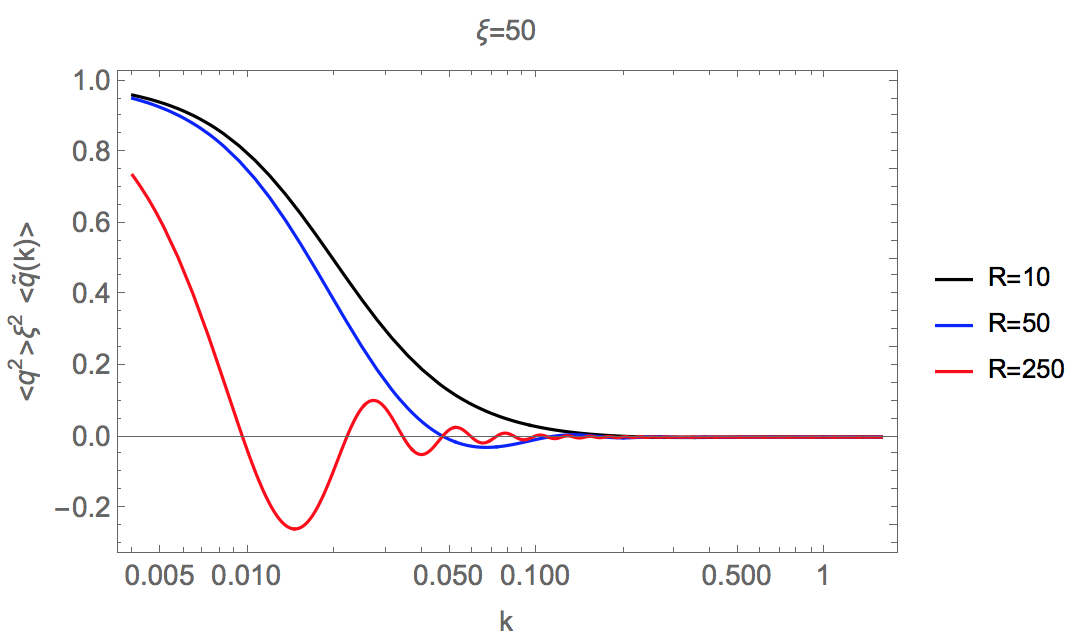}
\includegraphics[width=.5\columnwidth]{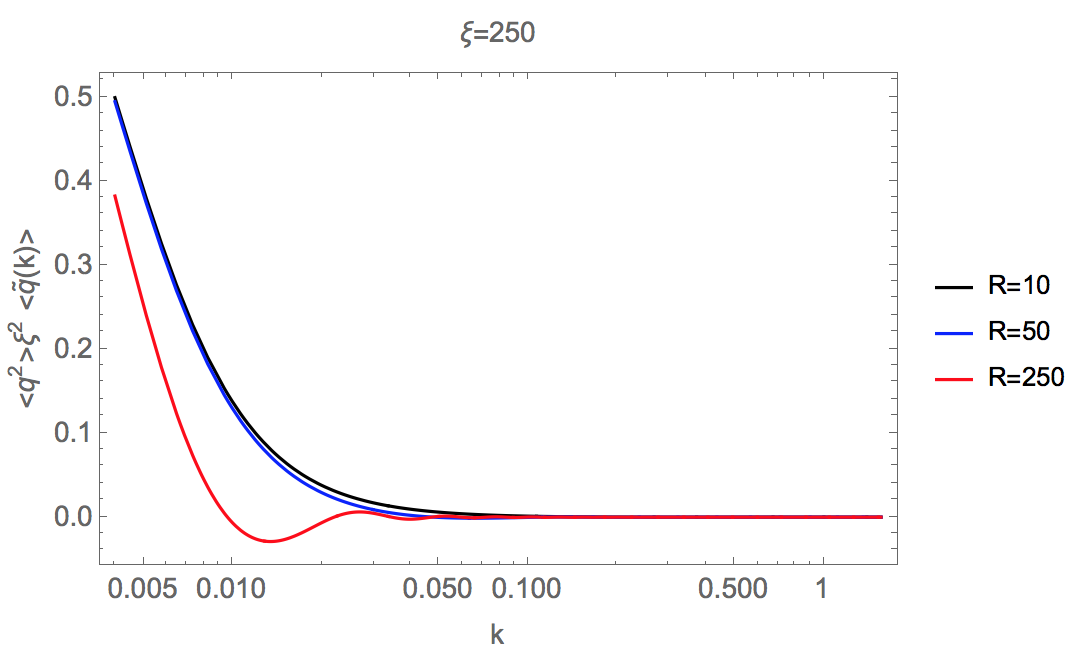}

\caption{Form factors of the  monopole charge elicited by fixed charges on the boundaries.}
\label{border}
\end{figure}

Finally, consider instead the possibility of a finite spin ice of radius $R$ in which boundaries are pinned as to impose a net charge to the system (we owe this idea to Andrew King of D-Wave Systems). If $R$ is large we can consider the continuum variable $x$ rather than $v$ and think of a charge $Q_p$ spread on a circular boundary as $\rho (x)=Q_p\delta(x-R)/(2\pi R)$ to which corresponds
\begin{equation}
\tilde \rho(k)=Q_p J_0(Rk) \simeq Q_p \sqrt{\frac{2}{\pi R q}}\cos(R q-\pi/4)~\text{for}~Rq\gg1 
\end{equation}
where $J_0$ is a modified Bessel function. Note that $~Rq\gg1$ means $x \ll R$ which is the region of interest: the bulk of the system. Then we have
\begin{equation}
\langle q(x) \rangle = \frac{1}{\langle q^2 \rangle} \int dy \langle q(x) q(y) \rangle \rho(y)
\label{pippo}
\end{equation}
from which we find that the following nonzero, oscillating form factor for the charge %
\begin{equation}
\xi^2_{||}\langle \tilde q(k) \rangle \simeq - \frac{Q_p}{\langle q^2 \rangle}  \sqrt{\frac{2}{\pi }}\frac{1}{1+\xi_{||}^2 k^2} \frac{\cos(Rk-\pi/4)}{\sqrt{R k}}
\label{ciccia}
\end{equation}
signals charge pinned on the boundaries.  We have used only the second term in Eq.~(\ref{pappa}), assuming the form factor experimentally recorded is that of the bulk.
We plot the form factor of Eq~(\ref{ciccia}) in Fig.~\ref{border}. 


Boundary charges are generally present in particle-based spin ice of finite size~\cite{nisoli2014dumping,nisoli2018unexpected,ortiz2019colloquium} where in fact $Q\propto R$. 

\begin{figure}[t!]
\includegraphics[width=.5\columnwidth]{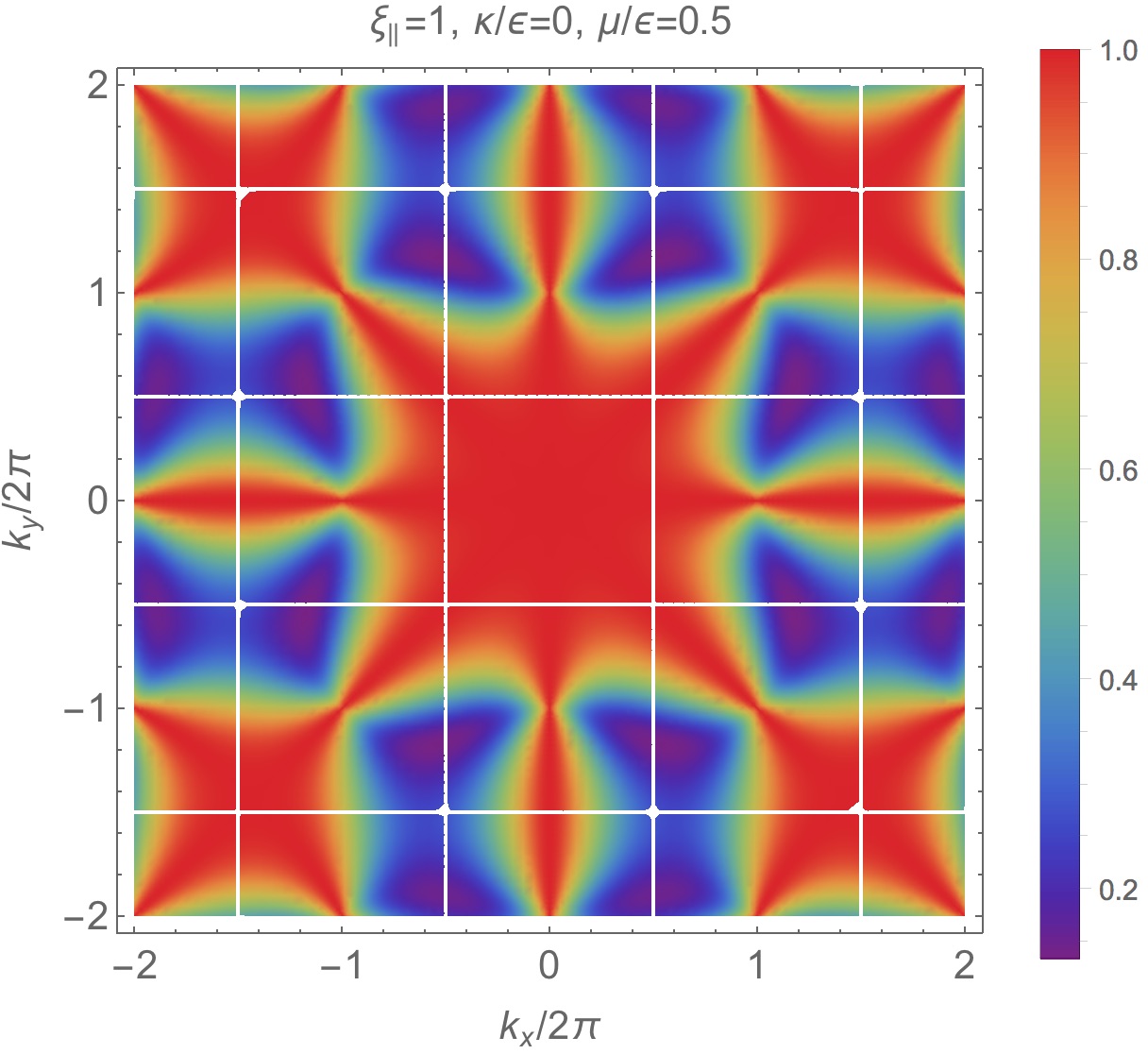}
\includegraphics[width=.5\columnwidth]{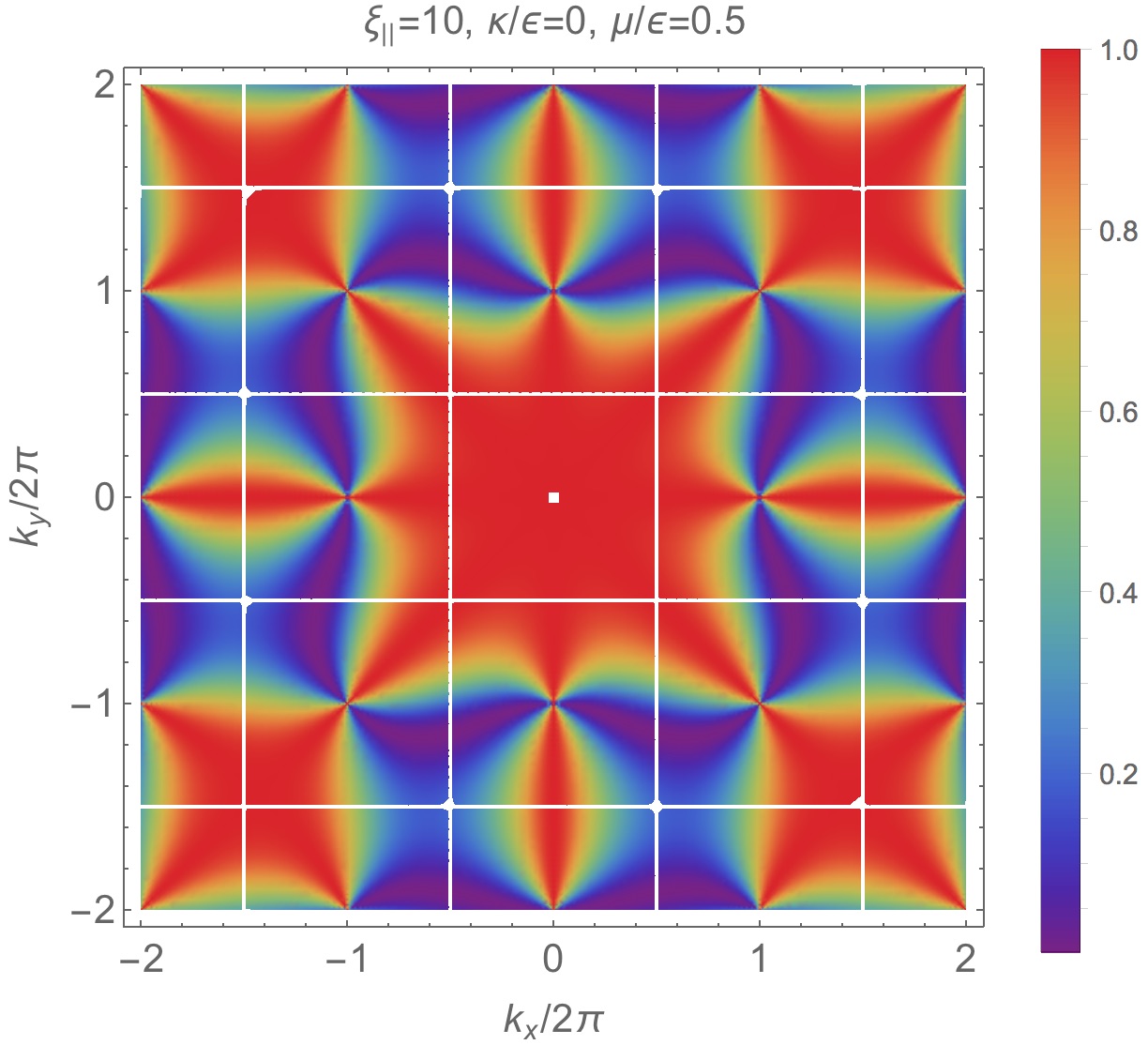}
\includegraphics[width=.5\columnwidth]{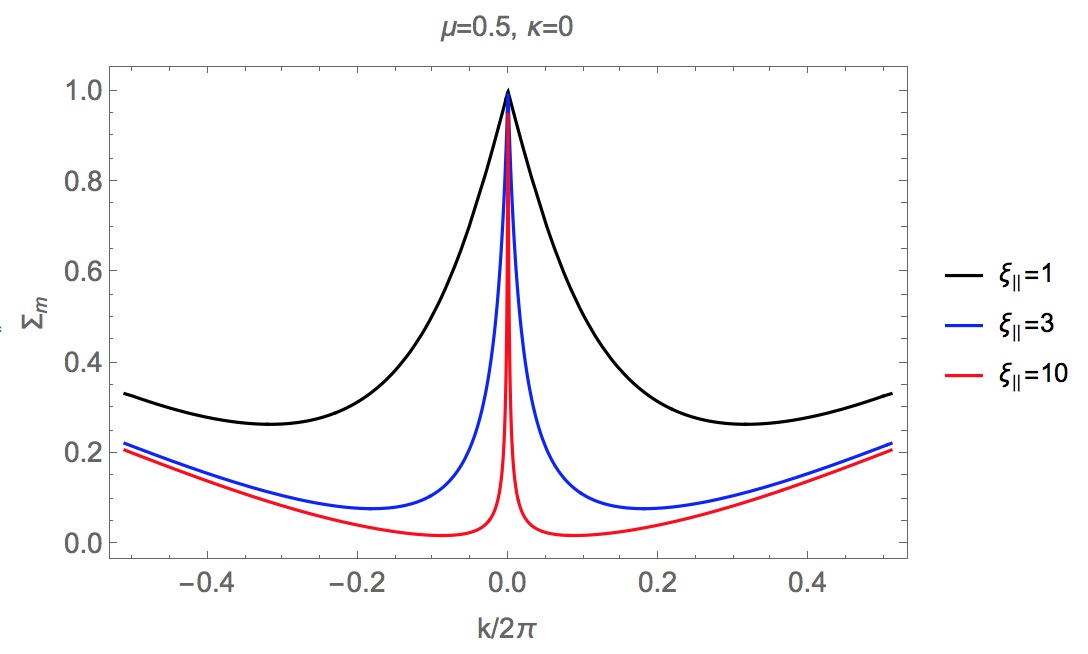}

\caption{Structure factors at high (top) and low (middle) temperature plotted from Eq.~(\ref{sigma}), for degenerate  square ice with monopole interaction. Notice how the monopole interaction sharpens the pinch points. Bottom: structure factor across a pinch point.}
\label{degenmu}
\end{figure}

\subsubsection{Algebraic Screening in Spin Ice with Monopole Interaction ($\epsilon>0, \mu>0,\kappa=0$)}

The  monopole-monopole interaction ($\mu>0$) introduces in 2D new physics not seen in 3D. 

As we know from our approach to general graphs, the entropic interaction is always screened Coulomb. 
Because in 3D the {\em real}  monopole-monopole interaction is also 3D-Coulomb, it does not alter the functional form of the screening, but merely changes  $ \xi_{||}=\sqrt{\epsilon/T}$  into $ \xi_{ ||}=\sqrt{\epsilon/(T+\mu)}$. 
Things are much different in 2D: the entropic interaction is a 2D-Coulomb, but the real interaction is a 3D-Coulomb. 

The first thing to notice is that we now have a new length-scale, the Bjerrum length $l_B=\mu/T$, above which thermal energy exceeds the strength of the monopole interaction. Not also that $l_B$ and $\xi_{||}$ scale differently with temperature and at high temperature $l_B<\xi$ while the opposite is true at low temperature. 

The next thing to notice is that $\tilde V(k)$ (the Fourier transform of $1/x$ on a 2D square lattice)  goes as $\tilde V(k) \sim 1/k$ at small $k$. 
Thus the denominator of the charge correlations in reciprocal space, which now take the form
\begin{equation}
\langle |\tilde q(k)|^2 \rangle= \frac{k^2}{1 + {\xi_{||}}^2 \gamma(\vec k)^2+ l_B  \gamma(\vec k)^2 \tilde V(k) },
\end{equation}
goes  linearly in $k$ as $\sim 1+ l_B k$ at small $k$ instead than quadratically as $1+ \xi_{||}^2k^2$ as it does when $\mu=0$. 
It can be seen clearly in the structure factor, which we plot in Fig.~\ref{degenmu}. In correspondence to the pinch points, the profile of the intensity is not a Lorenzian when $\mu\ne 0$ but has a weak singularity in the derivative, or $\Sigma_m\simeq 1-l_B |k|+(l_B^2-\xi_{||}^2)k^2$. In general, the visual effect is to make the structure factor sharper in correspondence to the pinch point at  $T\ne0$. 

More mathematically, this implies poles for $\langle |\tilde q(k)|^2 \rangle$ on the negative {\it real} axis in $k$ rather than on the {\em imaginary} axis and therefore lack of a screening length.

Consider a charge $Q_{\text{pin}}$ pinned in the system and   $\langle V_e(x)\rangle= \nabla^2 \langle q (x) \rangle$ the entropic potential generated by the screening charge $\langle q(x) \rangle$ elicited by $Q_{\text{pin}}$ [Eq.~(\ref{screenp})]. From Eqs~(\ref{taus}), (\ref{rhocorr}) we can write in the long wavelength approximation
\begin{equation}
\tilde V_e(k)=  \frac{T Q_{\text{pin}} }{i \epsilon \sqrt{\Delta} \langle  q^2 \rangle}\left(\frac{1}{k-k_+} -\frac{1}{k-k_-}\right)
\label{Ve}
\end{equation}
with
\begin{align}
2k_{\pm} &= \bar k \pm i \sqrt{\Delta} \nonumber \\
\bar k &=\mu/\epsilon \nonumber \\
\Delta&=4T/\epsilon -\mu^2/\epsilon^2.
\end{align}
Then, a frequently used trick is to write  $1/c=\int_0^{\infty}\! \exp(-z c) dz$ for a generic complex number $c$ with positive real part. Applying the trick to each term in Eq~(\ref{Ve}) and Fourier transforming one obtain the entropic potential elicited by the pinned charge as 
\begin{equation}
\langle V_e(x)  \rangle = \frac{ l_B }{2\pi \langle  q^2 \rangle} \int_{-\infty}^{+\infty} \! \! \! \frac{\lambda(z)}{(x^2+z^2)^{3/2}}dz
\label{Ve2}
\end{equation}
with 
\begin{equation}
\lambda(z)=Q_{\text{pin}} \frac{T^2}{\epsilon \mu} |z| \exp({-|z| \bar k/2}) \frac{\sin(|z|\sqrt{\Delta}/2)}{\sqrt{\Delta}}.
\label{lambda}
\end{equation}
for which  $\int \lambda(z) dz=Q_{\text{pin}}$. 
Thus the entropic potential  can be considered as generated by a  {\em virtual} density of charge spread on a line perpendicular to the plane (labeled by the coordinate $z$), intersecting the plane at the pinned charge, and of total charge $Q_{\text{pin}}$.

When $\Delta>0$, that is when $ T/\epsilon > \mu^2/4\epsilon^2$, the vertical, virtual charge distribution is confined by the a length $2/\bar k$. That length can be used for a ``multipole'' expansion of Eq.~(\ref{Ve}) in $x \bar k$. In particular, when $x \gg 2/\bar k= 2 \epsilon/\mu$ we have
\begin{equation}
\langle V_e(x)  \rangle \simeq \frac{ l_B }{2\pi \langle  q^2 \rangle} \frac{Q_{\text{pin}}}{x^3}
\label{Ve3}
\end{equation}
and by taking its Laplacian we obtain the charge density elicited by the pinned charge $Q_{\text{pin}}$ as
\begin{equation}
\langle q(x)  \rangle \simeq - \frac{ 9 l_B }{2\pi \langle  q^2 \rangle} \frac{Q_{\text{pin}}}{x^5}.
\label{rhopin}
\end{equation}

Instead, when $\Delta<0$, the sine in Eq~(\ref{lambda}) becomes hyperbolic. Now Eq.~(\ref{Ve3}), (\ref{rhopin})  are still valid but for $x \gg 1/  k_+ =l_B/2+\sqrt{l_B^2/4-1} \sim l_B$ when $l_B$ is large.

\begin{figure}[t!]
\includegraphics[width=.5\columnwidth]{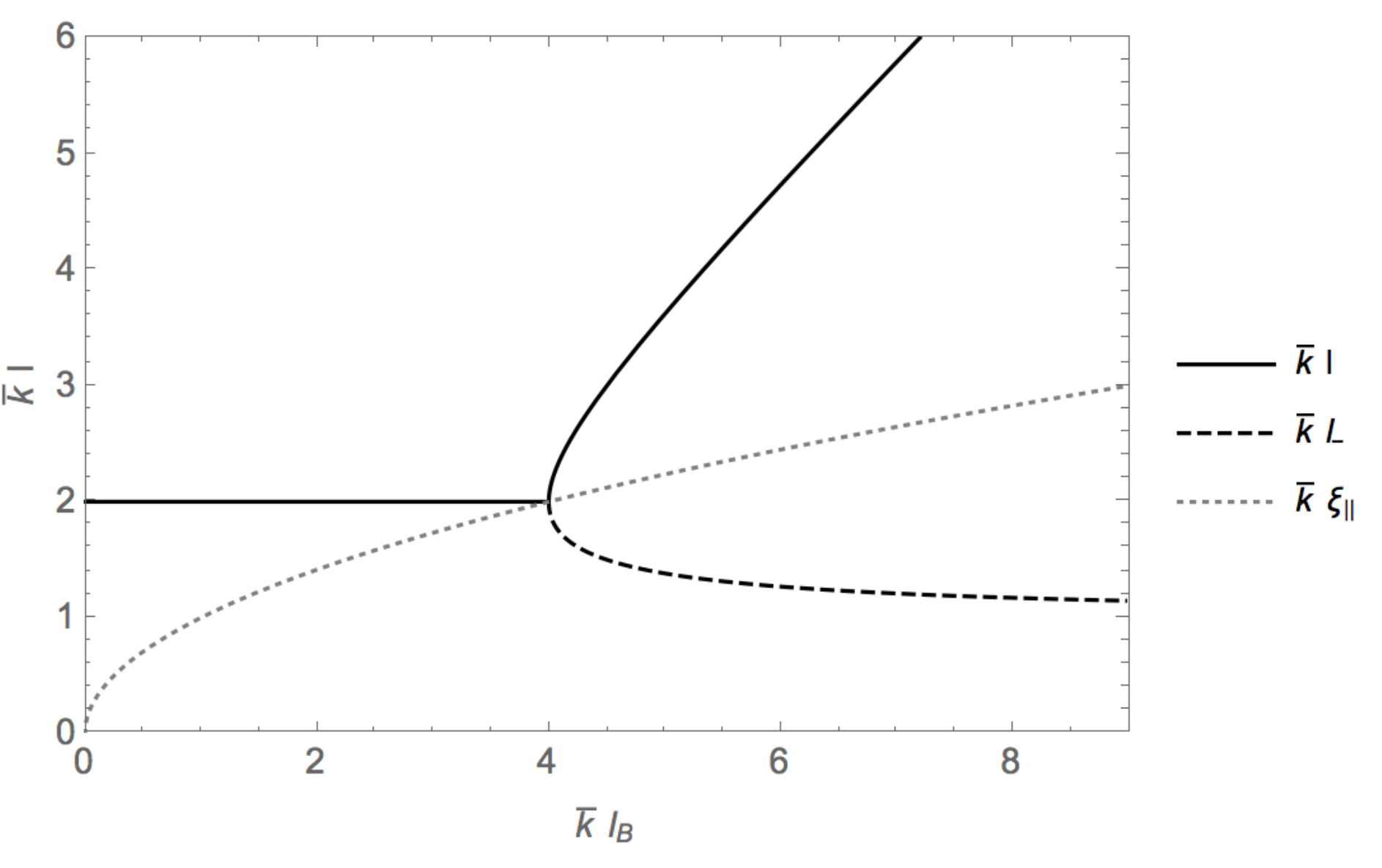}

\caption{Plot of $\bar k l$ (solid line) and $ \bar k l_-$ (dashed line) as a function of the normalized Bjerrum length $\bar k l_B$. Note that at low temperature both $\mu$ and $\epsilon$ should be renormalized and $\bar k =\mu/\epsilon$ might not remain constant. The region between the dotted lines is further split by the curve corresponding to the correlation length $\xi_{||}$, setting the length scale of the entropic interaction.}
\label{phase}
\end{figure}

Note that $\Delta>0$ means $\xi_{||}>l_B/2$. Thus, the difference between the two regimes depends on the interplay between the two relevant length scales, the screening length and the Bjerrum length. The two shapes (trigonometric or hyperbolic) for the screening charges thus denote two different regimes at small $x$, one of correlated charges and one of uncorrelated ones. For $\xi_{||}>l_B/2$, the presence of an imaginary part in the $k_{\pm}$ points to a {\em pseudo}-screening of length $2/\Delta \simeq \xi_{||} +l_B^2/8\xi_{||}$ for $ l_B/\xi_{||} \ll1$. That is, a short-length screening at small $x$, which then turns into the algebraic decay of the previous equations. For $\xi_{||}<l_B/2$ , one can interpret the elicited charge as coming from two linear distribution of virtual charges, one positive and screened along the $z$ axis on a longer length $ l= 1/ k_+$ and one negative and screened on a much smaller length $l_-=1/k_-$. The contribution of the second  charge can be taken as algebraic in the region $l_-<x<1/k_+$ whereas the full correlations become algebraic only for $x > 1/k_+$. There is therefore at small $T$ an interesting, intermediate region in spatial range that is worth studying (see Fig.~(\ref{phase}) and which corresponds to lengths lower than the Bjerrum length (since $l_+ \sim l_B$ at small $T$) and therefore strong monopole coupling. That region is further split in two by $\xi_{||}$ defining the length scale of the entropic interaction.  

We will study these regimes in more depth in the future. However, the most relevant feature here is that  in both regimes the elicited charge has an algebraic behavior at large $x$ in Eq~(\ref{rhopin}). Also, both requirements for the algebraic behavior imply $x \gg l_B$, as one would perhaps expect: the algebraic decay is valid at distances where the thermal fluctuations are stronger than the  monopole interaction.
The algebraic behavior at large distance implies that the monopole interaction impedes screening by keeping opposite charges close. And indeed, in studies of low-dimensional correlated electrons, it is well known that a 3D-coulomb interaction destroys screening in 2D~\cite{keldysh1979coulomb,jena2007enhancement,cudazzo2011dielectric}. 

In conclusion, square ice with $\mu\ne 0$ is an insulating rather than conducting plasma of charges, as lack of charge screening indicates. It becomes conducting when $\mu =0$. 

\subsection{Antiferromagnetically Ordering in Spin Ice ($\kappa<0$)}

  When $\kappa \ne 0$ from Eqs~(\ref{rhocorr}) we have
\begin{align}
\langle | \tilde  j(\vec {k}) |^2\rangle \simeq \frac{T}{\kappa}\left[1-\tilde \chi_{\!\perp}(k) \right].
\label{jcorrlw}
\end{align}
which relates correlations among currents to the perpendicular magnetic susceptibility.  

\begin{figure}[t!]
\includegraphics[width=.5\columnwidth]{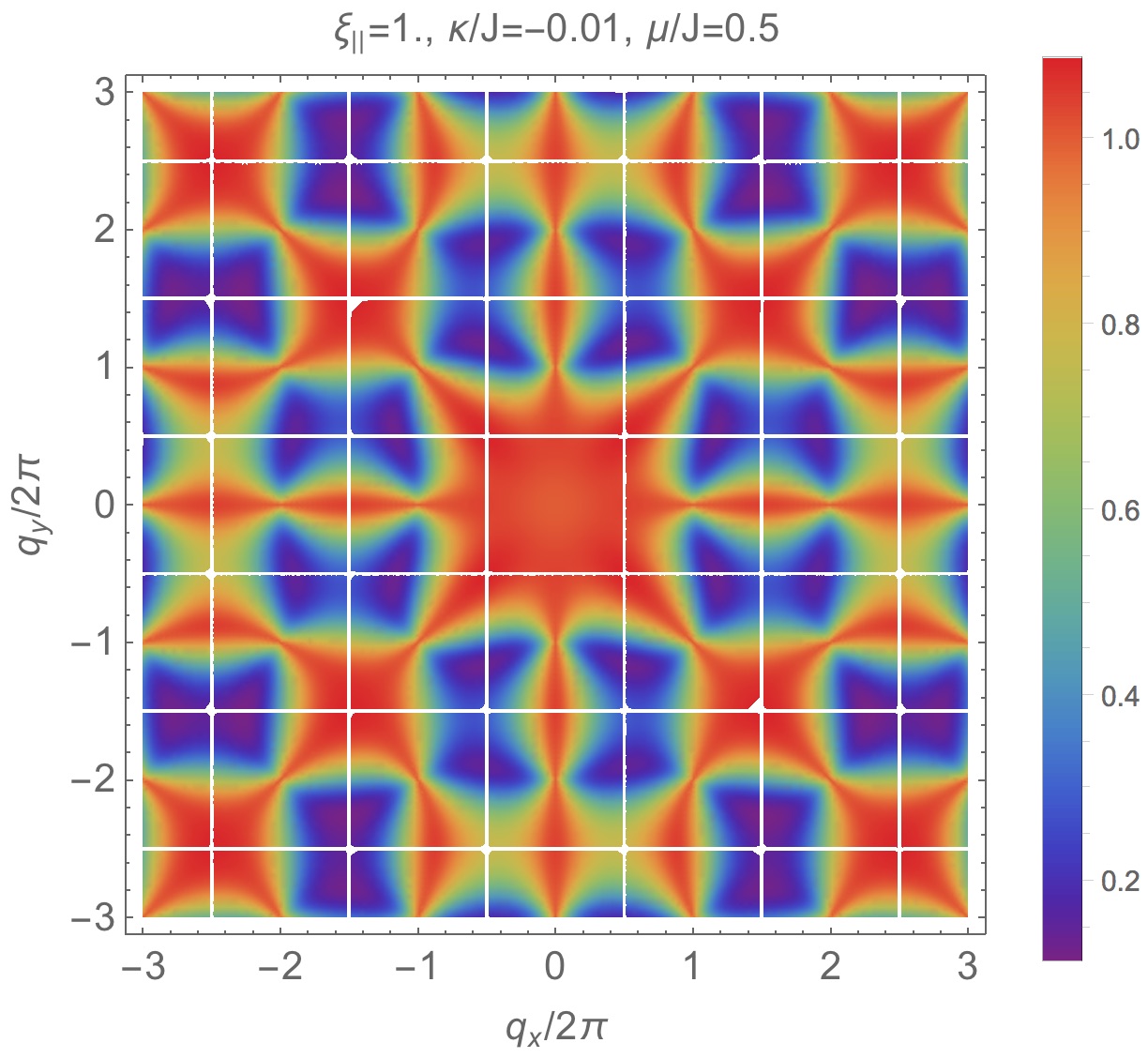}
\includegraphics[width=.5\columnwidth]{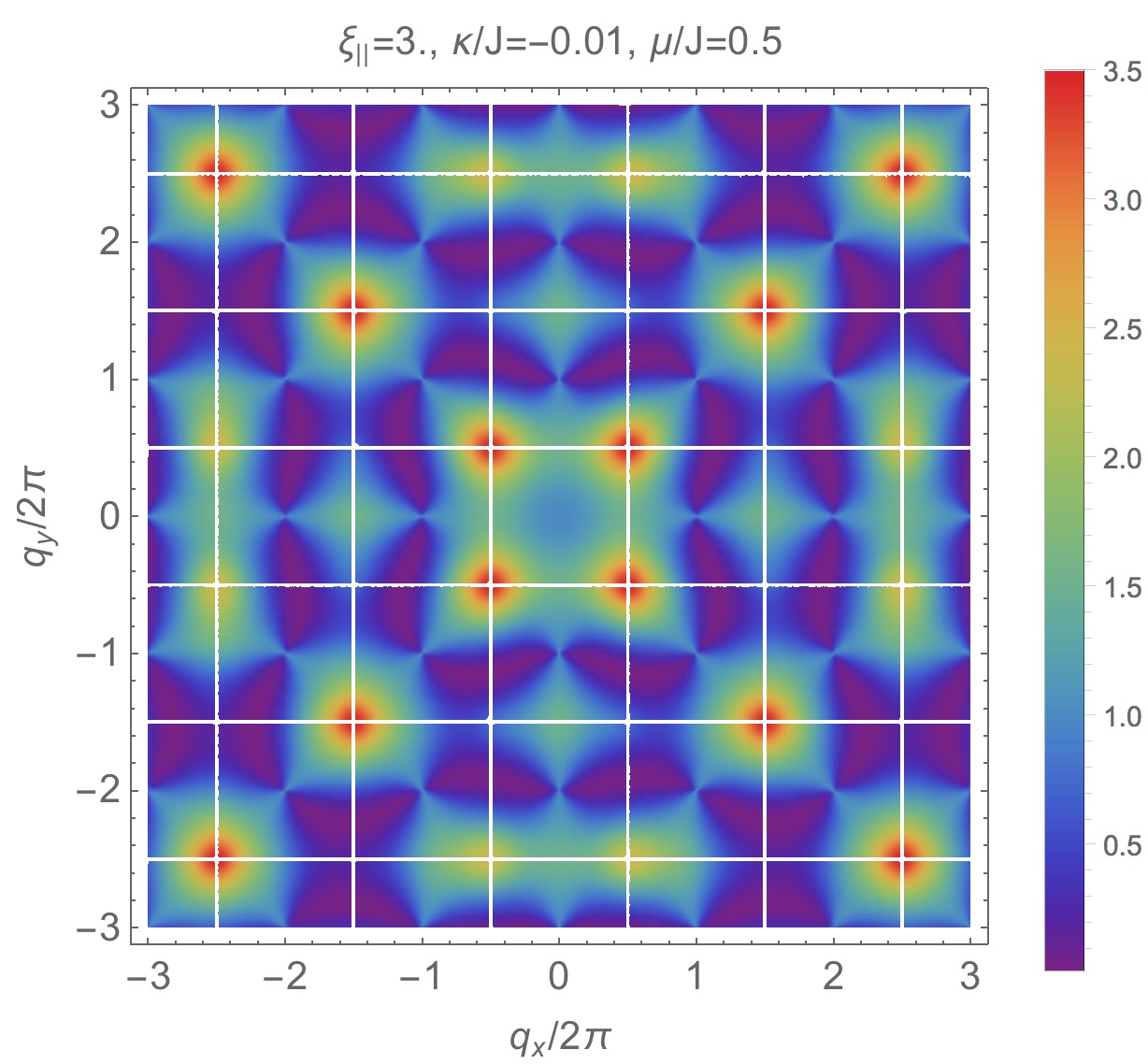}

\caption{Structure factors at different temperatures plotted from Eq.~(\ref{sigma}), for the FM square ice with monopole interaction.}
\label{AFM}
\end{figure}

When $k<0$,  ${\cal F}_2$ in Eq.~(\ref{f2-n}) is not bounded from below when $T\le T_c^{\text{afm}}= 8 |\kappa|$ (because $\gamma^2$ reaches its maximum $\gamma^2=8$ on the $K$ corners of the BZ). This  merely testifies of the expected ordering criticality in the AFM case. Currents are promoted by a negative $\kappa$ and the AFM state is an ice rule state that maximizes currents. There is a second order phase transition to such ordered state. 
At transition, our high $T$ approximation fails as interactions between fluctuations of the entropic fields cannot be neglected. Therefore, $T_c^{\text{afm}}$ is not the actual AFM  critical temperature, but merely a useful parameter in the context of our framework.

 For $T>T_c^{\text{afm}}$, $\tilde \chi_{\!\perp}(k)$ has a maximum on the $K$ corners of the BZ. 
Thus, we expand around $K$, $\gamma (K +\vec k)^2 \simeq 8- k^2$ in  $\tilde \chi_{\!\perp}(k)$ and from Eq~(\ref{jcorrlw}) we obtain for large $|p|$
\begin{align}
&\langle I_p I_{0}\rangle \simeq \frac{(T/\kappa)^2}{2\pi} (-1)^{p_x+p_y } K_0\left({|p|}/{\xi_{\text{afm}}}\right), 
\label{K0AFM}
\end{align} 
which expectedly alternates sign on adjacent plaquettes. The AFM correlation length is given by
\begin{equation}
\xi_{\text{afm}}^2=|\kappa|/(T-T_c^{\text{afm}})
\end{equation} 
 and is a measure of the size of the AFM domains. It diverges at the ordering criticality, though with the wrong exponent, as expected given our high $T$, quadratic approximation.  
 
Because $\kappa/\epsilon$ is small, $T^{\text{afm}}$ is much smaller than $2\epsilon$, the crossover temperature for the ice regime. Thus, above $T_c^{\text{AFM}}$  the divergence-free and divergence-full fields behave independently and features of the essentially transversal IM structure factor, such as pinch points and charge correlations, are still present. We plot the structure factor in Fig.~\ref{AFM}. Note there the maxima corresponding to the $K$ points of the BZ. These are routinely seen in experiments and often Monte Carlo simulations of degenerate square ice, that is of dipolar spin ice where the vertices are artificially made degenerate by vertical offsets of the islands. It is a clear indication that even though the vertices are degenerate the long range dipolar interaction wants to close fluxes. 

Note that also in this case the pinch points become sharper because of the presence of the monopole interaction.

\subsection{Ferromagnetic Case ($\epsilon>0, \kappa>0$)}

This case is rather interesting in itself, thought unlikely to be realized via current methods (see discussion at the beginning of this section). As we have discussed, it has an ordered, ferromagnetic ground state. Given the symmetry breaking, one would naively expect a second order phase transition. Of course, we do know of infinite order transitions  associated to symmetry breaking~\cite{lieb1967exact}, and we have already discussed how the ground state is the intersection of two topological requirements, the IM for $S$ and $^{\perp}\!S$ at the same time. While this does not describe at low temperature the realized cases of square ice with heigh offset exceeding the critical offset, which instead leads to a line liquid, it is a good approximation at intermediate temperatures. 

Our free energy at quadratic level suggests no critical temperature above zero. At the same time it suggests an exponential divergence in both correlation lengths $\xi_{\! \perp}, \xi_{||}$ as $T\to0$.

\begin{figure}[t!]
\includegraphics[width=.5\columnwidth]{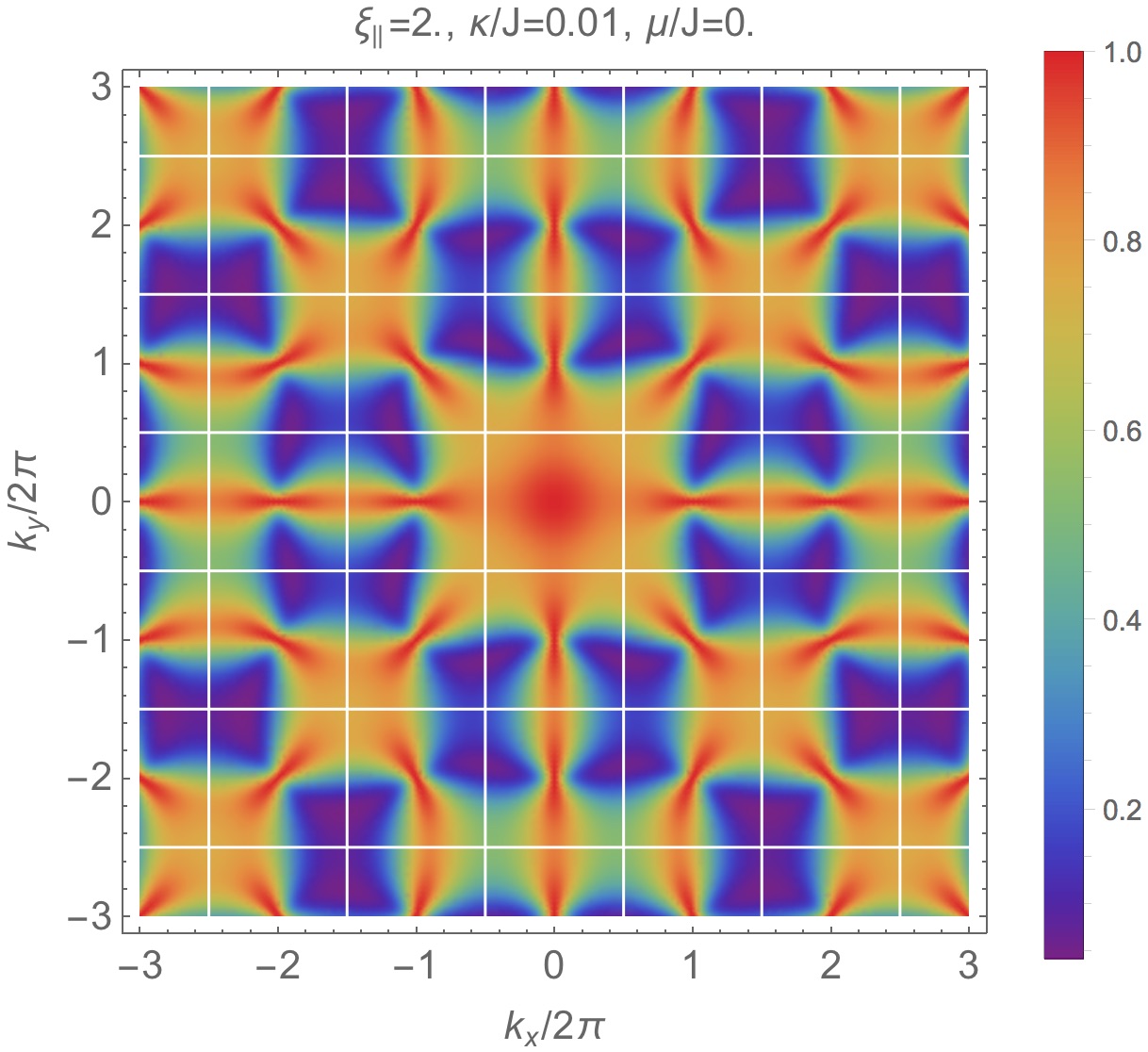}
\includegraphics[width=.5\columnwidth]{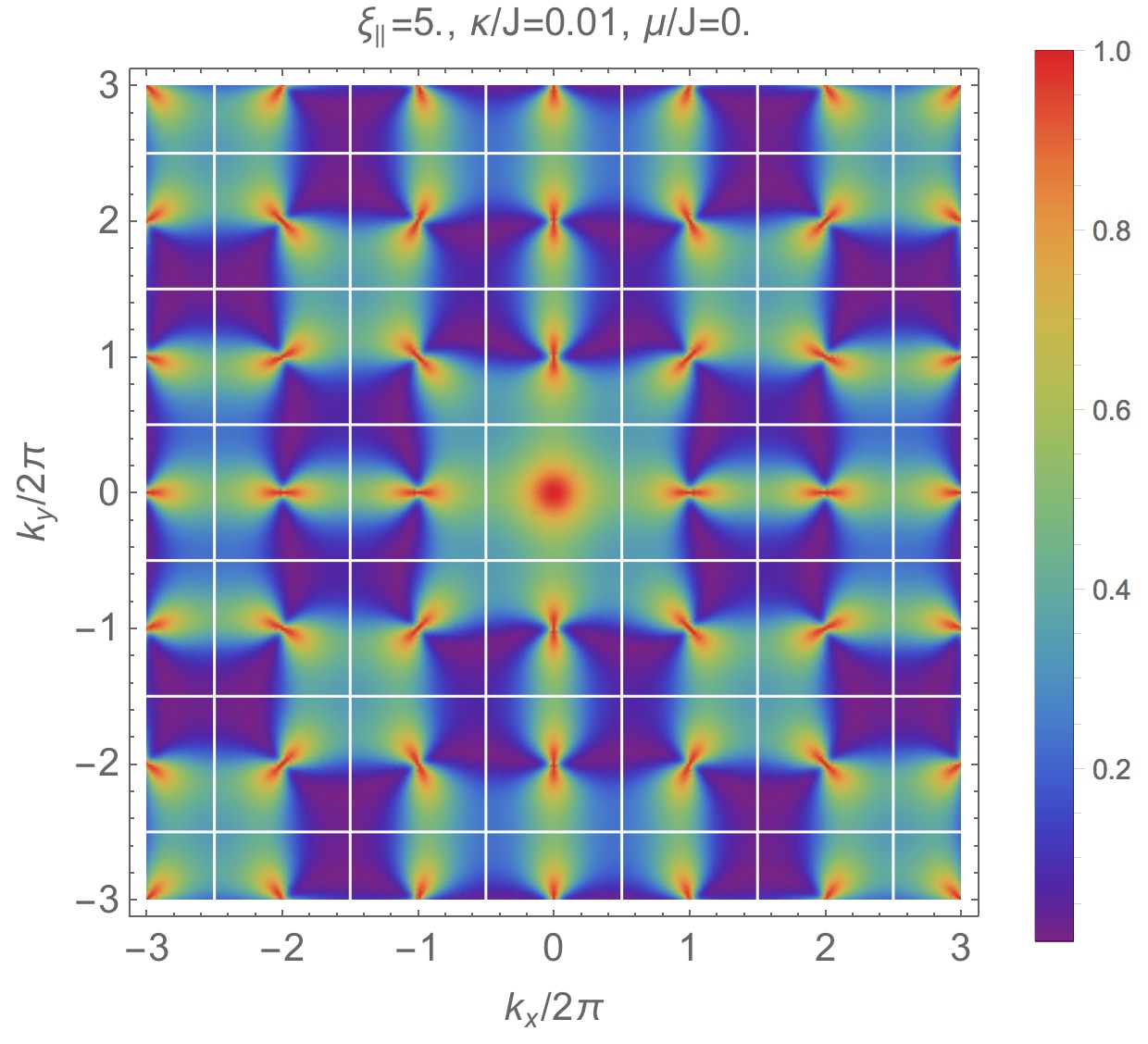}
\includegraphics[width=.5\columnwidth]{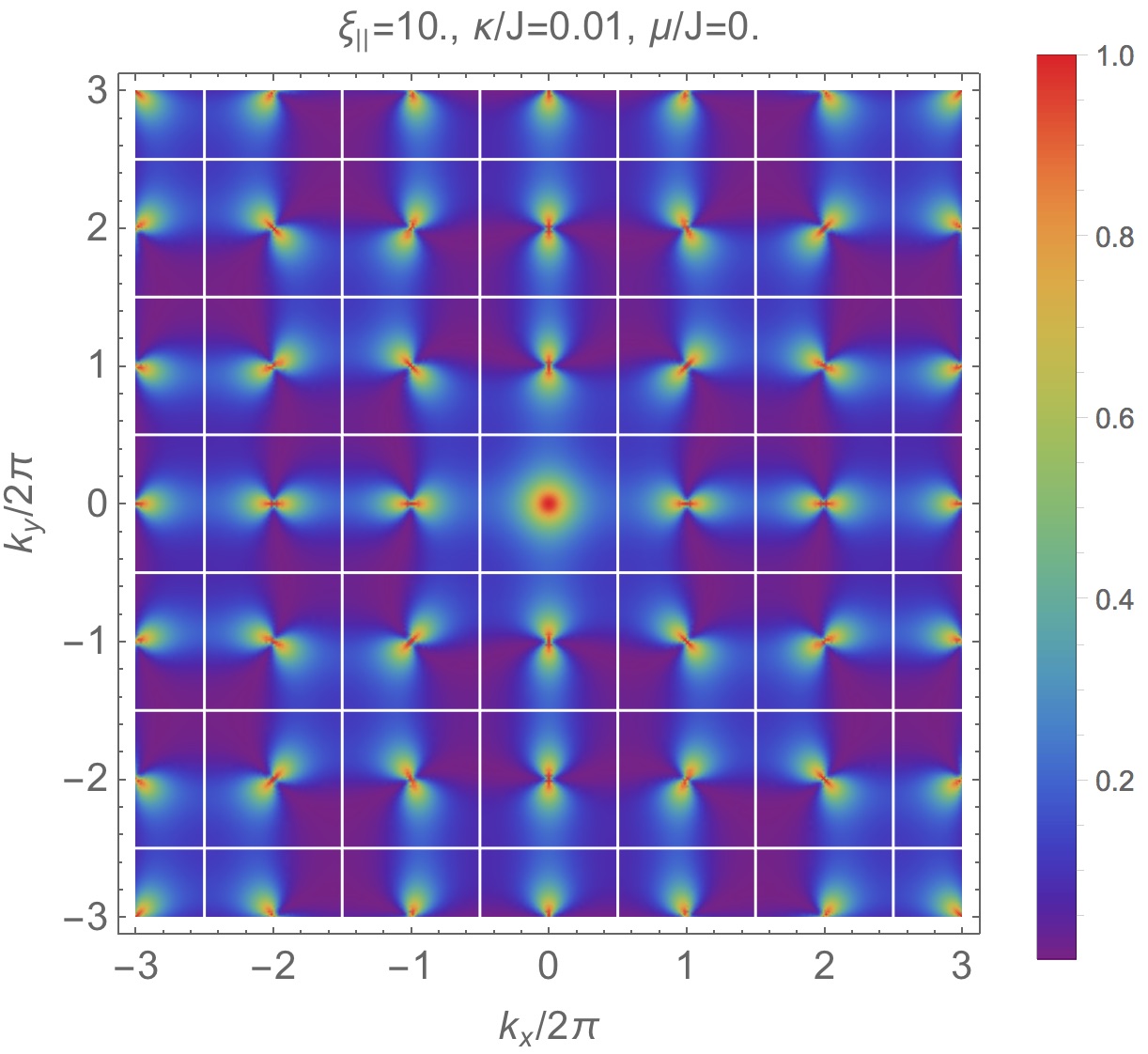}

\caption{Structure factors at different temperatures plotted from Eq.~(\ref{sigma}), for the FM square ice with monopole interaction.}
\label{FM}
\end{figure}

Given the gauge-free duality, everything said above for charge correlations applies here both to charges and currents. thus we have also screened correlations for currents as
\begin{align}
&\langle j_p j_{0}\rangle =  -\frac{(T/\kappa)^2}{2\pi} K_0\left({|p|}/{ \xi_{\! \perp}}\right). 
\label{K0}
\end{align} 
The structure factor (Fig.~\ref{FM})  is quite reminiscent of  spin ice. However, it has maxima at the pinch points, to signal incipient ferromagnetic ordering. Unlike the case $\kappa<0$, such maxima never diverge as temperature is reduced: they merely become sharper. 

We  conclude therefore that within our framework there is no phase transition and the system is critical only at exactly $T=0$, and correlation lengths diverge exponentially as zero temperature is reached. 

At temperatures higher than $\kappa$ but lower than $\epsilon$ we conclude that the system behaves as an excited spin ice but with screened 2D-Coulomb correlations among currents. When $\mu\ne0$ the discussion for the $\kappa=0$ applies and there is lack of screening among charges. 

\section{Conclusions}

We have proposed a general framework  to model spin ices on a general graph. We have then particularized it to square ice and studied it geometric specificity. 

We have obtained a series of results that are independent of geometry. We list them here.

\begin{itemize}

\item{The partition function of general Graph Spin Ice can be reformulated exactly as a functional integral over the charge distribution and its entropic field, the latter conveying the effect of the underlying spin ensemble [Eqs~(\ref{Z2})-(\ref{QFT3})].}

\item{The high $T$ behavior of a general Graph Spin Ice is described by a quadratic free energy of the average charges and contains informations of the graph via the graph Laplacian [Eq.~(\ref{QFT6})].}

\item{In absence of charge interaction and external fields and in the limit of high $T$, the entropic interaction among charges corresponds to the Green operator of the graph Laplacian. Thus, correlations are merely the {\it screened} Green operator of the graph Laplacian [Eqs~(\ref{qcorr5})-(\ref{qcorr1})]. Consequently, all topological properties of any unknown graph can be reconstructed from its spin ice behavior.}

\item{In absence of charge interaction and external fields and in the limit of high $T$,  the correlation length is $\xi^2=J/T$, a result already appreciated in various particular systems~\cite{garanin1999classical}.}

\end{itemize}

We have then particularized this treatment to the case of square ice. We list our main results

\begin{itemize}

\item{A general Hamiltonian can capture the gauge-free duality of the square geometry. By including a terms for currents one can better describe the threshold around pure spin ice. In practical realizations it is generally impossible to reach vertex degeneracy without also promoting currents.}

\item{The three cases, non surprisingly, have similar behaviors when $T$ is not too low, while remaining below the crossover to a spin ice state.}

\item{The formalism can be rather naturally ``pushed'' at low temperatures by properly renormalizing the monopole chemical potential $\epsilon$. When the ice manifolds is also a Coulomb phase, our formalism allows to conjecture with confidence and generality that $\xi\sim 1/ \langle q^2\rangle$ for $T\to 0$. This formula is exactly the generalization on a graph of the  Debye-H\"uckel screening length for a potential that is ``inverse-Laplacian'' and whose coupling constant is proportional to $T$, as our entropic potential is.  For a Coulomb phase charge is energetically gapped and  the mean square charge goes to zero exponentially. This leads to an exponential divergence in the correlation length which signals the topological nature of the Coulomb phase, and which  has been observed experimentally~\cite{fennell2009magnetic} in pyrochlores. We will show elsewhere that the low temperature relation between mean square charge and correlation function is maintained in bipartite graphs of odd coordination.  There, of course, $\langle q^2\rangle \to 1$ as $T\to0$ and there is no exponential divergence in the correlation length.}

\item{Monopole interaction in 2D has considerably different effects than in 3D. The concept is more general. On a graph, when the interaction is chosen such that $\hat V =\mu \hat L^{-1} $ little changes in the formalism, except a shift $T \to T+\mu$. This is precisely what happens in pyrochlore, where the entropic interaction is indeed found numerically to be the inverse of the laplacian, or $V(x) \propto 1/x$.}

\item{In 2D square ice we find that the monopole interaction destroys the entropic screening. Consequently, charge correlations are algebraic at non-zero temperature. Thus, square ice as a monopole plasma is  insulating at $\mu\ne0$ and conducting at $\mu =0$ .}

\end{itemize}

\section{Acknowledgements}
We thank  Andrew King (D-Wave Systems) for useful discussions and Beatrice Nisoli for proofreading.
This work was carried out under the auspices of the U.S.
DoE through the Los Alamos National
Laboratory, operated by
Triad National Security, LLC
(Contract No. 892333218NCA000001).

{\bf Data Availability Statement}

Data sharing is not applicable to this article as no new data were created or analyzed in this study.

\bibliography{library2.bib}{}

\end{document}